\UseRawInputEncoding
\documentclass[aps,prx,reprint,superscriptaddress]{revtex4-1} %preprint,showpacs,longbibliography,

\usepackage[colorlinks=true,linkcolor=blue,citecolor=blue,urlcolor=blue]{hyperref}

\usepackage{setspace} %to set 1.5 spacing in text
\usepackage{graphicx}
\usepackage{amsmath}
\usepackage{color}
\usepackage{amsmath}
\usepackage{amssymb}
\usepackage{verbatim}
\usepackage{latexsym}
\usepackage{enumerate} % to change the numbering of 'enumerate'
\usepackage{bm} % for bold face mathematical symbols
\usepackage{bbold}

\usepackage[caption=false]{subfig}
\usepackage{multirow}
\usepackage{booktabs}

\begin{document}

\title{Multi-gap topology and non-Abelian braiding of phonons from first principles}
% in monolayer Al$_2$O$_3$ via electrostatic doping %on the Brillouin zone boundary }

% \author[1,*]{Bo Peng}
% \author[2,*]{Adrien Bouhon}
% \author[1,${\dagger}$]{Robert-Jan Slager}
% \author[1,3,${\dagger}$]{Bartomeu Monserrat}

% \affil[1]{TCM Group, Cavendish Laboratory, University of Cambridge,
% J. J. Thomson Avenue, Cambridge CB3 0HE, United Kingdom}
% \affil[2]{Nordic Institute for Theoretical Physics (NORDITA),
% Stockholm, Sweden}
% \affil[3]{Department of Materials Science and Metallurgy, University of Cambridge, 27 Charles Babbage Road, Cambridge CB3 0FS, United Kingdom}
% \affil[*,$\dagger$]{Equal contribution}
% \affil[{}]{
% To whom correspondence should be addressed: bp432@cam.ac.uk (BP), adrien.bouhon@su.se (AB), bm418@cam.ac.uk (BM), rjs269@cam.ac.uk (RJS)}

\author{Bo Peng}
\thanks{equal author contribution}
%\email{bp432@cam.ac.uk}
\affiliation{TCM Group, Cavendish Laboratory, University of Cambridge, J. J. Thomson Avenue, Cambridge CB3 0HE, United Kingdom}
\author{Adrien Bouhon}
\thanks{equal author contribution}
%\email{adrien.bouhon@su.se}
\affiliation{Nordic Institute for Theoretical Physics (NORDITA), Stockholm, Sweden}
\author{Robert-Jan Slager}
\thanks{equal author contribution }
%\email{rjs269@cam.ac.uk}
\affiliation{TCM Group, Cavendish Laboratory, University of Cambridge, J. J. Thomson Avenue, Cambridge CB3 0HE, United Kingdom}
\author{Bartomeu Monserrat} 
\thanks{equal author contribution }
%\email{bm418@cam.ac.uk}
\affiliation{TCM Group, Cavendish Laboratory, University of Cambridge, J. J. Thomson Avenue, Cambridge CB3 0HE, United Kingdom}
\affiliation{Department of Materials Science and Metallurgy, University of Cambridge, 27 Charles Babbage Road, Cambridge CB3 0FS, United Kingdom}
%\thanks{$\textcolor{blue}*, \$$ equal author contribution}

%To whom correspondence should be addressed: bp432@cam.ac.uk (BP), adrien.bouhon@su.se (AB), bm418@cam.ac.uk (BM), rjs269@cam.ac.uk (RJS)}

\date{\today}

\begin{abstract}
Non-Abelian states of matter, in which the final state depends on the order of the interchanges of two quasiparticles, can encode information immune from environmental noise with the potential to provide a robust platform for topological quantum computation. We demonstrate that phonons can carry non-Abelian frame charges at the band crossing points of their frequency spectrum, and that external stimuli can drive their braiding. We present a general framework to understand the topological configurations of phonons from first principles calculations using a topological invariant called Euler class, and provide a complete analysis of phonon braiding by combining different topological configurations. Taking a well-known dielectric material, Al$_2$O$_3$, as a representative example, we demonstrate that electrostatic doping gives rise to phonon band inversions that can induce redistribution of the frame charges, leading to non-Abelian braiding of phonons. Our work provides a new quasiparticle platform for realizable non-Abelian braiding in {\it reciprocal space}, and expands the toolset for studying braiding processes.
%Double message: new material and the details of the braiding. Possibly the half patch invariants on the boundary? {\color{red} Shall we mention the ``half patch invariants on the boundary'' in the abstract?}
%AB:Suggestion (but maybe tricky, or not relevant): to propose double resonance Raman scattering protocol for the observation of braiding at the Brillouin zone boundary. Can anything be computed for that? Otherwise, do a qualitative argument. {\color{red} Not sure whether the electron and phonon wave vectors can induce double resonance Raman. I plotted it in Figure 11, but I am in doubt that it may be hard for experimentalists.}
\end{abstract}

%\flushbottom
\maketitle

%\begin{document}
%\flushbottom
%\maketitle

%\thispagestyle{empty}

\section{Introduction}

Non-Abelian states of matter arise from non-commutative interchanges of quasiparticles \cite{Stern2010}. In this process, called braiding, the winding of one quasiparticle around another can encode information, creating non-Abelian states that are immune from external noise: as long as the braiding occurs, the information is topologically protected. This process can form the basis for topological quantum bits (qubits), and a variety of strategies for non-Abelian braiding have been proposed, including braiding of quasiparticles in intrinsic topological states such as fractional quantum Hall systems \cite{DasSarma2005}, and also in symmetry-protected topological states such as half-quantum vortices in superconductors with $p$-wave symmetry \cite{DasSarma2006} and Majorana modes in hybrid systems \cite{Fu2008}. 
%Importantly, such schemes bring the concept of non-Abelian braiding of quasiparticles, usually inherent to intrinsic topological order such as FQHE systems, to exist in simpler systems such as topological band insulators, for example in the form of vortex modes. 
Multiple variants of these proposed architectures exist \cite{Iadecola2016,Ezawa2020}, and experimental evidence has started to emerge in the past few years \cite{Nadj-Perge2014,Wang2018}, providing a proof-of-principle for the existence of non-Abelian quasiparticles. However, multiple difficulties remain in exploiting these quasiparticles for braiding, for example Majorana fermions are boundary states that have proven challenging to observe \cite{Zhang2018c}. It would therefore be desirable to find alternative platforms in which non-Abelian braiding exists.

The recent development of the theory of topology in the energy bands of crystals \cite{Slager2013,Bernevig2006,Fu2011,Juricic2012,Hughes2011,Turner2012,Fang2012,Rainis2013,Slager2014,Shiozaki2014,Alexandradinata2014,Slager2015,Chiu2016,Shiozaki2017,Bouhon2017,Kruthoff2017,Po2017,Bradlyn2017,Rhim2018,Slager2019,Nakagawa2020,Uenal2019,Scheurer2020,Alexandradinata2020,Cornfeld2021,Heikkila2011,Xu2015,Lv2015,Yang2015,Fang2015,Huang2016b,Hu2016a,Yang2018,Armitage2018,Peng2018d,Peng2019,Diaz2019}
has created new opportunities for exploring non-Abelian braiding of band crossing points (nodes) in reciprocal space \cite{Bouhon2020a,Ahn2019a,Ahn2019b,Bouhon2020,Tiwari2020,Uenal2020}, providing an alternative to the real space braiding exploited by other strategies. Real space braiding is practically constrained to boundary states, which has made experimental observation and manipulation difficult; instead, reciprocal space braiding occurs in the bulk states of the band structures and we demonstrate in this work that this provides a straightforward platform for non-Abelian braiding. Concretely, reciprocal space braiding occurs between nodes carrying non-Abelian frame charges in multi-band systems described by a real Hamiltonian \cite{Bouhon2019,Wu2019a,Ahn2019a,Ahn2019b,Bouhon2020,Uenal2020,Tiwari2020,Jiang2021,Bouhon2021,Lange2021,Peng2021,Chen2021,Koenye2021}, and as such is referred to as \textit{multi-gap topology}, in contrast to the single-gap topology associated with two-band systems. The real Hamiltonian constraint leads to a real basis of eigenvectors \cite{Zhao2017b,Wieder2018,Wu2019a,Ahn2019a,Ahn2019b,Bouhon2019,Bouhon2020,Tiwari2020,Uenal2020} and is fulfilled with the symmetry requirement of either (i) a combination of $C_2$ rotation symmetry and time reversal symmetry $T$ for both spinful and spinless systems, or (ii) a combination of spatial inversion symmetry $P$ and time reversal symmetry $T$ for spinless systems. In any multi-band system arising from a real Hamiltonian, the band crossing points carry non-Abelian frame charges that can be converted through the braiding of nodes belonging to adjacent energy gaps \cite{Wu2019a,Bouhon2019,Tiwari2020,Beekman2017}. If a pair of nodes within a gap carry the same frame charge (with the same sign), they cannot be annihilated when brought together. Conversely, two nodes within the same gap with opposite frame charges can be annihilated. When nodes are braided, the signs of their non-Abelian frame charges flip, thus changing the relative stability between pairs of nodes in the same gap. As a consequence, the braiding of nodes is accompanied with the transfer of stable pairs of nodes from one gap to an adjacent gap.

To achieve elementary braiding it is necessary to braid one node of an energy gap with a node of an adjacent gap (the gap immediately above or immediately below in energy). The motion of the nodes can be driven by modifying the band structure of the material with the application of external stimuli. In solid state systems, where the band structure inherits the symmetries of the crystal, braiding involves groups of nodes that are related by symmetry. Moreover, the braid trajectories usually collapse onto the high-symmetry points of the Brillouin zone. As a consequence, the transfer of a stable pair of nodes from one gap to an adjacent gap is often manifested by a band inversion at one of the high-symmetry points, and such band inversion provides a very direct signature of the braiding process in crystalline systems. This phenomenology is rather general, and reciprocal space braiding can in principle occur in multi-band systems of \textit{any} quasiparticle, including electrons and phonons.

%To achieve braiding it is necessary to wind the nodes around one an-other, which can be accomplished by modifying the band structure with the application of external stimuli. Such braiding processes can flip the signs of the frame charges,and if a pair of nodes carry the same frame charge (with the same sign), they cannot be annihilated when brought together. 

In this work we argue that phonons are an ideal platform to study the non-Abelian braiding of band nodes in the context of multi-gap topology. As a bosonic excitation associated with ionic vibrations, the entire phonon spectrum is readily accessible to external probes. This contrasts with fermionic excitations such as electrons, in which only band nodes near the Fermi level can be accessed, placing significant restrictions for the full exploitation of multi-gap topologies. Additionally, phonons are charge neutral, spinless quasiparticles, and time reversal symmetry $T$ is hard to break in phonons because they do not directly couple to magnetic fields. For these reasons, the symmetry requirements of real Hamiltonians can be easily fulfilled by a wide range of materials, suggesting that many will exhibit non-Abelian frame charges in their phonon dispersion. This motivates us to extend the study of phonon bands from single-gap topologies, which have been extensively studied \cite{Stenull2016,Zhang2018a,Miao2018,Li2018a,Xia2019,Zhang2019c,Liu2020,Li2021,Peng2020a,Liu2020b,Wang2021,Tang2021,Liu2021,Liu2021a,Liu2021b,You2021,Xie2021,Zheng2021,Wang2021a}, to multi-gap topologies, which remain largely unexplored.

%not until recently have we proposed the existence of topological phonons with non-Abelian frame charges, formed by at least three bands, in the context of the so called multi-gap topologies \cite{}.

The main purpose of our work is to provide a general framework to study non-Abelian braiding of phonons using first principles methods, enabling the accurate material-specific calculation of multi-gap topologies in the phonon bands of any material. The key mathematical objects to study multi-gap topologies are (i) non-Abelian frame charges, and (ii) a topological invariant called \textit{Euler class}, which captures the relative stability of a pair of nodes within the same gap, that is, whether annihilation of the nodes is possible or not, and that in turn depends on their trajectory with respect to the nodes of the adjacent gaps. We describe the calculation of the Euler class for phonon bands using the phonon eigenvectors that can be obtained from a first principles lattice dynamics calculation. We also explain how to then use the Euler class to derive the global topological configurations of all the nodes in the multi-band system. % in the two neighboring gaps 

%that can be extended to other quasiparticles. We summarize the theoretical background to derive non-Abelian frame charges in a three-band system with $C_2T$ symmetry, and introduce the concept of a new topological invariant called patch Euler class to understand the relative stability of a pair of nodes, i.e., whether a pair of nodes can be annihilated or not. We develop computational techniques that can compute the patch Euler class for phonons using their eigenvectors obtained from lattice dynamics calculations, and use the patch Euler class to understand the global topological configurations of all the nodes in the three-band system.

To illustrate our method, we study non-Abelian braiding of phonons in monolayer Al$_2$O$_3$ from first principles.
%, we employ first principles calculations, and provide a new material candidate with new strategies to control the braiding. 
We show that Al$_2$O$_3$, a well-known dielectric material, carries non-Abelian frame charges in its phonon dispersion, and that braiding within a three-band subspace can be driven with electrostatic doping.
%, the phonon bands are inverted, redistributing the nodes with non-Abelian frame charges and leading to novel braiding processes. 
We explain in detail how to determine the Euler class to construct consistent topological configurations during the braiding process, thus providing a template for analogous calculations in other materials. Our main prediction -- non-Abelian braiding constrained by crystal symmetry and driven by electrostatic doping in monolayer Al$_2$O$_3$ -- constitutes a robust proposal for the practical realization of this phenomenon. Additionally, Al$_2$O$_3$ has been widely incorporated into electronic devices as a dielectric material \cite{Gilmer2002,Hirama2012,Ren2017,Kwon2020}, and as a result, electrostatic doping using a gate voltage could be seamlessly integrated into the modern microelectronics industry. 
%The focus on Al$_2$O$_3$ offers new opportunities for studying braiding-related phenomena in this otherwise well studied material. Given the versatility of Al$_2$O$_3$ and its varied uses in technological applications, our results on its phonon braiding properties can greatly expand its applicability as a material platform.

The paper is structured as follows. In Sec.\,\ref{theory}, we review the theoretical background behind non-Abelian frame charges, the associated patch Euler class, and the Dirac strings that connect nodes with half-integer Euler class \cite{Ahn2019a}. We then present a general computational methodology to calculate the Euler class of phonon band crossing points from first principles in Section~\ref{computing}. In Section~\ref{case}, we apply this methodology to study non-Abelian braiding of phonons driven by electrostatic doping in monolayer Al$_2$O$_3$ taking place in a three-band subspace. In Section~\ref{discussion}, we present our conclusions and discuss future research directions.

\section{Theoretical background}
\label{theory}

\subsection{Non-Abelian frame charges}
We start by introducing non-Abelian frame charges in the context of a three-band Bloch Hamiltonian of a two-dimensional system with $C_2T$ symmetry. %as an example. %, and use the case with
%In the following, we use an example of a 2D system with $C_2T$ symmetry. 
The spectral decomposition of the $3\times 3$ Hamiltonian gives 
\begin{equation}
 H=\sum_{n=1,2,3}\vert e_n\rangle E_n \langle e_n \vert,   
\end{equation}
with the ordered eigenvalues $E_1 < E_2 <E_3$ (which we assume to be gapped) and the eigenvectors $\{ \vert e_{n} \rangle \}_{n=1,2,3}$ that can be chosen to be real in an appropriate basis~\cite{Bouhon2020}. As a result, the three real and normalized eigenvectors form a three-dimensional orthonormal frame ($\vert e_{1} \rangle,\vert e_{2} \rangle,\vert e_{3} \rangle$) $\in$ $\mathbb{R}^3\times \mathbb{R}^3$, that is, an orthogonal matrix $\mathsf{O}(3)$, or, fixing the handedness (i.e.~the orientation of the frame), a 3D rotation matrix $\mathsf{SO}(3)$. For real eigenvectors, the gauge phase degree of freedom of complex eigenvectors turns into a $+/-$ sign degree of freedom
%; that is, 
%(from the phase freedom of complex eigenvectors) 
\cite{Wu2019a,Bouhon2020}; specifically, ($\vert e_{1} \rangle,\vert e_{2} \rangle,\vert e_{3} \rangle$), ($\vert e_{1} \rangle,-\vert e_{2} \rangle,-\vert e_{3} \rangle$), ($-\vert e_{1} \rangle,\vert e_{2} \rangle,-\vert e_{3} \rangle$), and ($-\vert e_{1} \rangle,-\vert e_{2} \rangle,\vert e_{3} \rangle$) all represent the same state (the orientation of the frame is not preserved if only one sign flips). % because only the magnitude of the eigenvector matters.
Therefore, the order-parameter space of the Hamiltonian can be expressed as $\mathsf{SO}(3)$ modulo the group of $\pi$ rotations that flip the sign of two eigenvectors, namely $\mathsf{SO}(3)$/$D_2$ (the dihedral point group $D_2=\{E,C_{2},C'_{2},C''_{2}\}$ is composed of three independent and perpendicular $\pi$-rotations) \cite{Beekman2017}.

For a band crossing point (node) in a three-band system, we can define a topological frame charge by the geometry of the $\mathsf{SO}(3)$ rotations encircling the node in momentum space, as the acquired angle can be calculated by decomposing the 3D rotation matrix around the encircled node \cite{Wu2019a,Bouhon2020,Johansson2004,Tiwari2020}. A closed path in $\mathsf{SO}(3)$/$D_2$ can be characterized by the fundamental homotopy group $\pi_1$[$\mathsf{SO}(3)$/$D_2$] = $\mathbb{Q}$, where $\mathbb{Q} = (\pm i, \pm j, \pm k, -1, +1)$ is the quaternion group with $i^2=j^2=k^2=-1$, $ij=k$, $jk=i$, $ki=j$, and where the charges $\{i,j,k\}$ anti-commute \cite{Wu2019a}. Lifting the $\mathsf{SO}(3)$-frame in the covering spin group $\mathsf{Spin}(3)=\mathsf{SU}(2)$ and using the correspondence $(-{\rm i}\sigma_x,-{\rm i}\sigma_y,-{\rm i}\sigma_z )\mapsto (i,k,j)$ from the parallel-transported spin-frames over a base loop $l$ and the quaternion charges, we can assign a quaternion frame charge to any node formed by the bands within the region bounded by the loop $l$~\cite{Wu2019a}. The nodes formed by the lower two bands can thus be characterized by the frame charge $\pm i$, and the nodes formed by the upper two bands can be characterized by $\pm j$. A pair of nodes, one formed by the lower two bands and the other formed by the upper two bands, can be denoted by $\pm k$ as $k=ij$. We note that the sign of the charges is fixed by $(i)$ the gauge chosen for the frame at a base point of the loop, and $(ii)$ the orientation of the parallel-transport over the base loop. The stability of a pair of nodes formed by the same two bands can then be characterized by $-1$, because two nodes with the same frame charge cannot annihilate when brought together, as can be deduced from the algebra of the quaternion charges, i.e.~$i^2=j^2=k^2=-1$ (note $k^2=i\cdot j\cdot i \cdot j=i\cdot j\cdot (-j) \cdot i=i^2=-1$). We also conclude that the $-1$ frame charge is gauge invariant since $(+i)^2=(-i)^2$, and similarly for $j$ and $k$. On the other hand, two nodes with opposite frame charges can be annihilated or created pairwise, with a total frame charge of $(-i)(+i)=(-j)(+j)=+1$. %\textcolor{magenta}{We emphasize that the relative stability of a pair of nodes from the same gap is gauge invariant as it describes two different physical processes, i.e.~either the nodes .} 

%and the band inversion (transferring nodes between the upper and the lower bands) can be interpreted as switch between $i^2=-1$ and $j^2=-1$.

\begin{figure}
\centering
\includegraphics[width=0.9\linewidth]{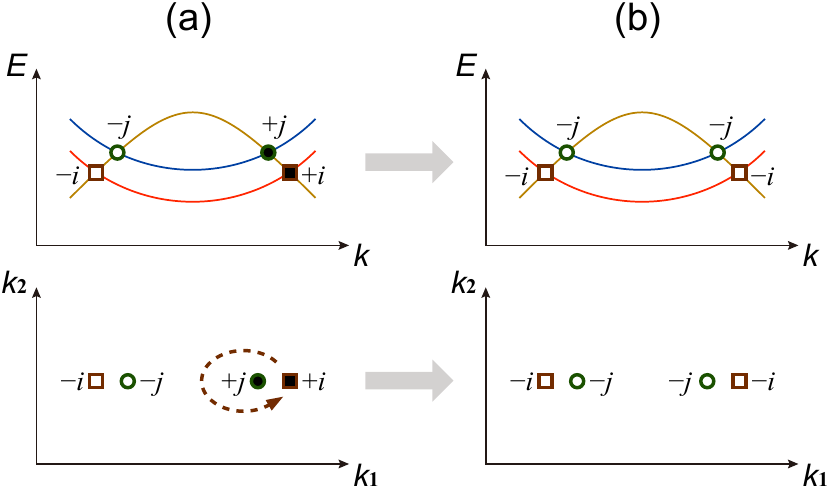}
\caption{Topological configurations (a) before and (b) after the braiding. Squares (circles) represent the nodes formed by the lower (upper) two bands, and open (closed) symbols represent the nodes with negative (positive) frame charges. Importantly, the frame charges of the nodes are unambiguously defined only once a base point, an oriented base loop and choice of gauge have been fixed. We implicitly assume that these have been fixed and we do not show them in the figure.
}
\label{schematic} 
\end{figure}

The quaternion group is non-Abelian, as captured by the non-commutativity of the charges $\{i,j,k\}$ and the non-trivial action of their conjugation, e.g.~$j^{-1}ij =-jij= j^2i=-i$. This indicates the possibility of flipping the non-Abelian frame charge of a node. Keeping in mind that the frame charges are defined for a fixed base point with a fixed oriented base loop, we can easily see, through the composition of oriented loops, that a conjugation operation, say $j^{-1}ij=-i$, corresponds to the braiding of a node, $i$, around a node in an adjacent gap, $j$ \cite{Wu2019a}. Therefore, the band nodes in three-band systems can carry a non-Abelian charge, and the sign of the topological frame charge can be flipped by a braiding processes, as schematically shown in Fig.~\ref{schematic}. Hereafter we use open (closed) symbols to represent the nodes with negative (positive) frame charges. Different from the topological invariants formed only by two bands in single-gap topologies, in three-band systems the frame charge depends on the braiding of the nodes formed by both the lower two and the upper two bands, which involve all three bands \cite{Bouhon2020a}. For clarity, we use squares (circles) to represent the nodes formed by the lower (upper) two bands. As shown in Fig.~\ref{schematic}(a), a node formed by the lower two bands with a frame charge $+i$ (labelled as a closed square) can circle another node formed by the upper two bands with a frame charge $+j$ (labelled as a closed circle), and as a result, the signs of the frame charges are changed to be $-i$ and $-j$ respectively [labelled as an open square and an open circle in Fig.~\ref{schematic}(b)]. With this process we can create an obstruction to annihilate two opposite nodes, e.g.~$+i$ and $-i$ in Fig.~\ref{schematic}(a), by braiding one node around another node formed by the neighboring bands, so the frame charges of the pair of nodes become the same, e.g.~the two nodes with the same frame charge of $-i$ in Fig.~\ref{schematic}(b). 

While the quaternion charges intrinsically unveil the non-Abelian nature of the multi-gap topology of systems described by a real Hamiltonian, they are cumbersome to use in real material band structures because crystal point group symmetries lead to node multiplicities. Fortunately, there exists a complementary quasi-two-dimensional topological invariant which not only simplifies the computation of the topological charges, but also greatly refines the characterization of the topological stability of nodes belonging to the same gap. This is the patch Euler class that we introduce next.

%For the two nodes formed by the same bands (either $\pm i$ or $\pm j$), whether they can annihilate depends on the signs of the frame charges: The band nodes with the opposite frame charges annihilate when brought together, whereas the pair with the same frame charge cannot annihilate.

%{\color{red} I am not sure whether the frame charges $+i/-i/+j/-j$ have the sign freedom too? If so we should mention it here...}

\subsection{Patch Euler class}

In the following, we number the bands from lower to higher energy with $E_n \leq E_{n+1}$, and we label the partial gap between two successive bands $n$ and $(n+1)$ as $\{\Delta_n\}$ with $E_n\leq \Delta_n \leq E_{n+1}$. Similar to the Berry curvature in single-gap topologies, we can compute the Euler curvature (Euler form) for bands $n$ and $(n+1)$ in gap $\Delta_n$ \cite{Peng2021,Ahn2019a,Bouhon2020}:
\begin{equation}
    {\rm Eu}_n (\textbf{k}) =  \langle \partial_{k_1} e_{n} \vert \partial_{k_2} e_{n+1} \rangle - \langle \partial_{k_2} e_{n} \vert \partial_{k_1} e_{n+1} \rangle ,
    \label{Euler}
\end{equation}
where $\vert e_{n} \rangle$ and $\vert e_{n+1} \rangle$ are eigenvectors of band $n$ and ($n+1$) respectively, and $\textbf{k}=(k_1,k_2)$ are the coordinates of the Brillouin zone. The Euler class $\chi_n$ for the bands $n$ and $n+1$ over a patch $\mathcal{D}$ of the Brillouin zone (assuming that there is no node connecting the bands $n$ and $n+1$ to other bands on $\mathcal{D}$) is then defined by \cite{Bouhon2020,Ahn2019a}
\begin{equation}\label{Eq:patchEuler}
\chi_n[\mathcal{D}] = \frac{1}{2\pi}  \left[ \int_{\mathcal{D}} {\rm Eu}_n (\textbf{k}) dk_1 dk_2 - \oint_{\partial\mathcal{D}} \textbf{{\rm a}}_n \cdot d\textbf{k} \right],
\end{equation}
where $\partial \mathcal{D}$ is the boundary of the patch, and with the Euler connection ${\rm a}_{n,i} = \langle e_n \vert \partial_{k_i} \vert e_{n+1} \rangle $ for $i=1,2$.
%where $k_a$ and $k_b$ are sampled in the patch for the Euler-form integral.

\begin{figure}[h]
\centering
\includegraphics[width=0.9\linewidth]{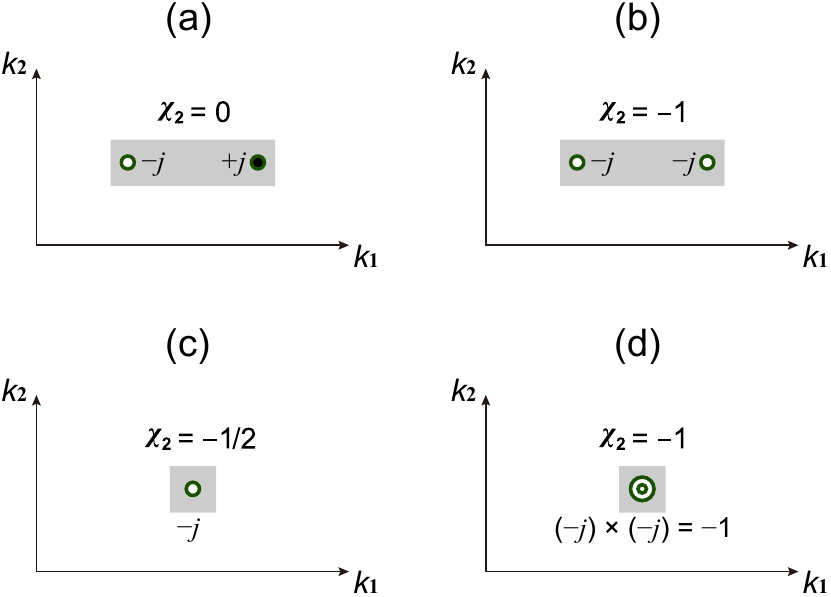}
\caption{Patch Euler class for two nodes in the same gap $\Delta_2$ (a) with the opposite frame charges and (b) with the same frame charges, as well as the patch Euler class for (c) a linear node and (d) a quadratic node. We use open (closed) symbols to represent the nodes with negative (positive) frame charges, and one symbol (two concentric symbols) to represent the linear (quadratic) node.} 
\label{patchEuler} 
\end{figure}

When integrated over the whole Brillouin zone, the Euler class $\chi_n$ is an integer $\mathbb{Z}$ \cite{Ahn2018,Ahn2019a,Xie2020a,Bouhon2019,Bouhon2020a}, and indicates the presence of $| \chi_n |$ pairs of stable nodes formed by a two-band subspace. When integrated over a patch $\mathcal{D}$, the Euler class in Eq.\,(\ref{Eq:patchEuler}) can either have integer or half-integer values, indicating the presence of $2| \chi_n |$ stable nodes within the patch. For instance, assuming the presence of two nodes within the patch, an Euler class of 0 indicates that the nodes can annihilate [Fig.~\ref{patchEuler}(a)], whereas an Euler class of $\pm 1$ means that they cannot annihilate when brought together [Fig.~\ref{patchEuler}(b)]. Importantly, the Euler class is related to the frame charges discussed above: for a patch containing only one node, a patch Euler class of $\chi_1 = \pm 1/2$ ($\chi_2 = \pm 1/2$) can be associated to the frame charge $\pm i$ ($\pm j$) and such node is referred to as a linear node [Fig.~\ref{patchEuler}(c)], while the value $\vert \chi_n \vert = 1$ indicates the frame charge $-1$ and such node is referred to as a quadratic node [Fig.~\ref{patchEuler}(d)]. Hereafter, we label a quadratic node of $\vert\chi_n\vert=1$ as two concentric symbols since it can be interpreted as the superposition of two linear nodes. 

We note that the Euler class refines the topological analysis since, contrary to the frame charges, it keeps track of the stability of an arbitrary number of nodes formed by two bands, taking a half-integer (integer) value for an odd (even) number of stable nodes. However, we also note that this requires that the two bands under consideration must be disconnected from all the other bands over the patch. The direct computation of the frame charges remains useful when more than two bands are degenerate at a single point, which can happen at critical points during a band inversion (or in systems with three-dimensional irreducible representations protected by cubic point groups).

%But before we discuss Dirac strings, 
One interesting feature of the Euler class is that it gives the lower bound of the power-like dispersion of the bands at a band crossing \cite{Jiang2021,Peng2021}. More precisely, %when all the nodes of the patch lie on top of each other (e.g.~at a high-symmetry point of the Brillouin zone where two- or three-dimensional irreducible representations appear, or when an accidental band crossing takes place due to a band inversion), 
the number $2\vert \chi_n\vert$ gives the lower bound of the exponent of the leading term in a Taylor expansion of the energy eigenvalues at the band crossing. %Hereafter, we label a double node of charge $\vert\chi_n\vert=1$, i.e.~exhibiting a minimally quadratic dispersion in all directions, as two concentric symbols since it can be interpreted as the superposition of two linear nodes. 
At this stage, a clarification is necessary to distinguish between electrons and phonons. In electronic band structures, the dispersion at band crossings almost always realizes the lower bound indicated by their Euler class because of the strong electrostatic screening provided by electrons. By contrast, and as described below, the frequencies of the phonon bands correspond to the square root of the eigenvalues of the dynamical matrix that defines the topology. 
%($\omega \sim \sqrt{\lambda}$). 
Interestingly, the order of band crossings in phonon band structures is almost always doubled, with the exception of the dispersion of the acoustic bands at $\Gamma$ corresponding to the Glodstone modes of the system~\cite{Lange2021}.
% E.g.~the nodes with a linear dispersion in frequency a quadratic dispersion for the dynamical matrix eigenvalues ($\lambda\sim \omega^2$), i.e.~with a patch Euler class $\vert\chi_n\vert=1/2$ must have a minimum dispersion  of the frequency corresponds to a quadratic dispersion of the eigenvalues.

Because the gauge sign ($\pm 1$) of the real eigenvectors is not fixed, the absolute sign of the topological charges is not uniquely defined. For example, if we flip the gauge signs of the orthonormal frame of eigenvectors from ($\vert e_{1} \rangle,\vert e_{2} \rangle,\vert e_{3} \rangle$) to ($\vert e_{1} \rangle,-\vert e_{2} \rangle,-\vert e_{3} \rangle$), the sign of the patch Euler class $\chi_2$ also flips, and similarly for the frame charges. Therefore, the sign of the Euler class and of the non-Abelian frame charge is gauge dependent, and for an individual node taken in isolation this sign has no physical meaning. However, the relative sign between two distinct nodes is not gauge dependent as it provides information on the stability of the nodes. Therefore, we can compute the Euler class for different patches, and assign the sign of the topological frame charges to get a consistent global topological configuration where the relative signs of all nodes agree with their gauge invariant relative stability. Such a global picture can be obtained by fixing the gauge globally \cite{Peng2021}. For this, we introduce in Sec.\,\ref{subsec:dirac} the last conceptual object that is needed, namely the Dirac string \cite{Ahn2019a}.

\subsection{Dirac strings} \label{subsec:dirac}

If we consider two linear nodes in gap $\Delta_n$, each band eigenvector forming the nodes $\{\vert e_n\rangle , \vert e_{n+1}\rangle \}$ carries a $\pi$ Berry phase on any loop encircling one node at the time, thus indicating a $\pi$ disinclination line connecting the two nodes. The gauge sign of the eigenvectors must then flip when crossing this line \cite{Ahn2018}. We can use a Dirac string to visualize the line connecting the two nodes, in analogy with the Dirac string connecting two Weyl nodes in 3D indicating the winding of the $\mathsf{U}(1)$ gauge phase of complex eigenvectors.

The trajectory of the Dirac string is not unique because we can change the gauge signs of the eigenvectors (although such local change does not affect the topological stability of whether any pair of nodes can be annihilated when merged). Despite the fact that we can assign different Dirac strings for the same pair of nodes, the trajectory of a Dirac string between two nodes is constrained by the gauge freedoms of all other nodes, collectively leading to what are known as the ``Dirac string rules'' \cite{Ahn2019a,Jiang2021}:
\begin{enumerate}
\item All linear nodes formed by the same two bands must be connected by Dirac strings in pairs, whereas the quadratic nodes can be interpreted as two linear nodes merged together with an internal Dirac string.
\item The sign of the frame charge of a node in $\Delta_n$ changes when crossing a Dirac string connecting two nodes in the neighboring gaps ($\Delta_{n-1}$ or $\Delta_{n+1}$). This can be realized either by moving the node in $\Delta_{n}$ across a fixed Dirac string in $\Delta_{n-1}$ or $\Delta_{n+1}$ [Fig.~\ref{Dirac}(a)] or by moving a Dirac string in the neighboring gap across the fixed node [Fig.~\ref{Dirac}(b)]. 

\begin{figure}[h]
\centering
\includegraphics[width=0.9\linewidth]{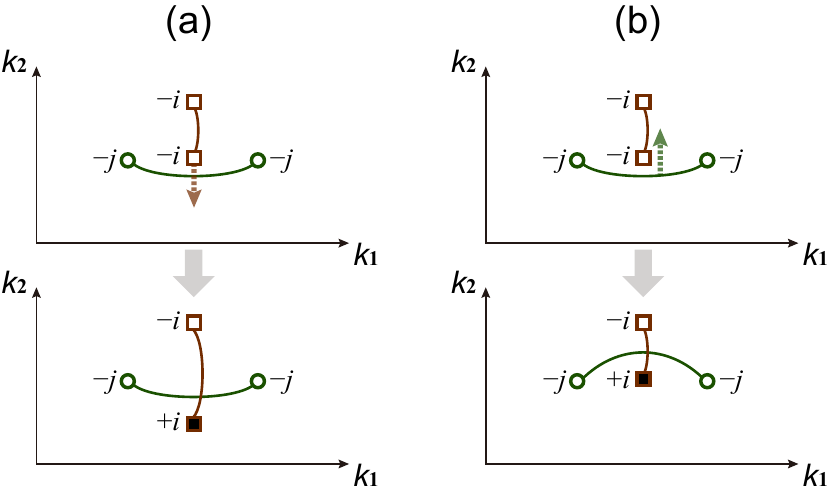}
\caption{Flipping of the sign of the frame charge by (a) moving the node across a fixed Dirac string or (b) moving a Dirac string across the fixed node. We note that these processes amount to the same thing.}
\label{Dirac} 
\end{figure}

\item All the Dirac strings connecting the nodes in the same gap can be re-assigned by changing their start and end nodes. For example, in Fig.~\ref{recombine} we can connect node 1 with node 3 and node 2 with node 4, or connect node 1 with node 2 and node 3 with node 4, or connect node 1 with node 4 and node 2 with node 3.

\begin{figure}[h]
\centering
\includegraphics[width=0.9\linewidth]{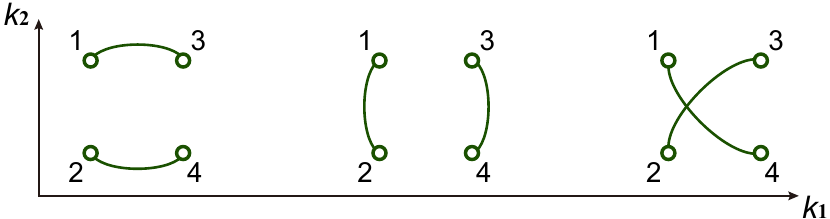}
\caption{Dirac strings re-assigned by changing the start and end nodes.}
\label{recombine} 
\end{figure}

\end{enumerate}

By systematically computing the Euler class of every single band crossing, and then of every pair of band crossings in the same gap, we can assign a signed frame charge to every node, as well as the Dirac strings that connect every pair of linear nodes, such that we obtain a consistent global topological configuration \cite{Jiang2021,Peng2021}, as discussed in the next subsection. This procedure works like a puzzle: we arbitrarily fix the sign, i.e.~the gauge, of an initial node and then we iteratively assign the Dirac strings of the neighboring nodes in a consistent manner, i.e.~under the constraint of the gauge invariant values of all the patch Euler classes previously computed.

\subsection{Global topological configuration}

To obtain the global topological configuration, we start by dividing the Brillouin zone into different patches, each containing a pair of nodes in the same gap, %Each patch overlaps with neighboring patches 
and the patches cover all the nodes. We then calculate the Euler class for each patch, which provides information on the relative stability within each and every pair of the nodes.

We then specify the relative signs of the frame charges for all the nodes based on the patch Euler class calculations, and afterwards connect all the linear nodes in pairs by Dirac strings to make the Euler class for all the patches consistent with each other. Even within each patch there are two possible configurations: a patch Euler class of $\pm 1$ (0) can either correspond to two nodes with the same (opposite) frame charge(s), or contain two opposite (same) charge nodes and an extra Dirac string in the neighboring gap. %Similarly, a patch Euler class of $\pm$ 1 can either contain two nodes with the same frame charge, or indicate two nodes with the opposite frame charges 

We next check whether the assignment of frame charges and Dirac strings is physically consistent for the global topological configuration. As long as the frame charges and the Dirac string of one patch are fixed, the rest of the global topological configuration can be deduced like a puzzle based on the computed Euler class for all the patches and the Dirac string rules.

Because the local gauge sign is not fixed, the start and end nodes of the Dirac strings, as well as the corresponding trajectory, are not unique. Therefore, for a given set of patch Euler classes, there may be many different (but consistent) global topological configurations. The difference comes from the local and global gauge choices. However, the different global topological configurations capture the same physics, for example, whether a pair of nodes will annihilate or not, which is gauge invariant. 

We can apply the same strategy, based on the calculated Euler class and the Dirac string rules, to obtain the topological configurations during the braiding processes that take place when the system undergoes a transformation of its band structure, i.e.~during a topological phase transition upon band inversion. However, once the global topological configuration of the system is known, the topological configuration of any other phases reached upon the displacement of the nodes and band inversions can be readily predicted by applying the conversion rules presented above. 

In real materials, where the crystal symmetries constrain the movement of the nodes and often collapse the braid trajectories to the high-symmetry points, these rules need to be complemented with the crystal symmetry rules contained in the irreducible representations which dictate when a band crossing can be avoided or not. These concepts are exemplified in Sec.\,\ref{case}, which describes the braiding of phonons in aluminium oxide. 

Overall, this establishes a theoretical framework to study non-Abelian braiding of \textit{any quasiparticle} with three bands in their spectra, as long as the system has $C_2T$ symmetry so that the corresponding Hamiltonian is real \cite{Bouhon2020}. We refer the reader to Refs.~\cite{Ahn2018,Wu2019a,Bouhon2020,Bouhon2020a} and especially Ref.~\cite{Peng2021} for more details. The theoretical formalism we have described can also be extended to spinless systems of any dimensions with $PT$ symmetry.

\section{Computational methodology}
\label{computing}

The key quantity in the theoretical formalism described in the previous section is the Euler form in Eq.\,(\ref{Euler}). Its evaluation requires the eigenvectors of the quasiparticles as input, and these eigenvectors can be directly calculated from first principles for a range of quasiparticles. In this work, we use phonons as an example quasiparticle to describe how to calculate the patch Euler class. As discussed earlier, we focus on phonons because (i) their bosonic nature means the entire spectrum is accessible (in contrast to the restriction to the Fermi energy in fermionic systems), and (ii) the time reversal symmetry $T$ is hard to break in phononic systems.

\subsection{Lattice dynamics}

The ions in solids vibrate around their equilibrium positions \textbf{r}($l \kappa$) with displacements \textbf{u}($l \kappa$), where $l$ and $\kappa$ label the unit cells and the atoms in each unit cell respectively. Under the harmonic approximation, the potential energy surface can be expressed as a quadratic function of the displacements of the atoms \cite{Born1954,Togo2015}
\begin{equation}
\Phi = \Phi_0 + \frac{1}{2}\sum_{ll'\kappa\kappa'}\sum_{\alpha\beta}
\Phi_{\alpha\beta}(l\kappa,l'\kappa')
%\frac{\partial ^2 \Phi_{\alpha\beta}(l\kappa,l'\kappa')}{\partial u_{\alpha}(l\kappa) \partial u_{\beta}(l'\kappa')}
u_{\alpha}(l\kappa) u_{\beta}(l'\kappa'),
\end{equation}
where $\alpha$ and $\beta$ are the Cartesian indices, $\Phi_0$ is the zeroth order force constant evaluated at the ionic equilibrium positions, and $\Phi_{\alpha\beta}(l\kappa,l'\kappa')$ are the second order force constants which can be evaluated as the second derivatives of $\Phi$ with respect to ionic displacements, or equivalently as the first derivatives of the atomic force on atom $l'\kappa'$ under an atomic displacement \textbf{u}($l \kappa$) \cite{Togo2008,Togo2015}
\begin{equation}
\Phi_{\alpha\beta}(l\kappa,l'\kappa') =
\frac{\partial ^2 \Phi}{\partial u_{\alpha}(l\kappa) \partial u_{\beta}(l'\kappa')} = -\frac{\partial F_{\beta}(l'\kappa')}{\partial u_{\alpha}(l\kappa)}.
\end{equation}
This can be calculated using the finite differences method \cite{Kunc1982,Alfe2009,Monserrat2018} or using density functional perturbation theory \cite{DFPT}.

The dynamical properties of the ionic motion are then determined by the dynamical matrix $D(\textbf{q})$, which plays the role of the Hamiltonian, and is obtained from the second order force constants as 
\cite{Dove1993,Ziman,Srivastava1990}:
\begin{equation}
   D_{\kappa\kappa'}^{\alpha\beta}(\textbf{q}) = \sum_{l'}\frac{\Phi_{\alpha\beta}(0\kappa,l'\kappa')}{\sqrt{m_{\kappa}m_{\kappa'}}}
   \textrm{e} ^{i\textbf{q}\cdot [\textbf{r}(l'\kappa') - \textbf{r}(0\kappa)]},
\end{equation}
where \textbf{q} is the wave vector and $m_{\kappa}$ is the mass of atom $\kappa$.
%calculating the eigenvalues and eigenvectors of the dynamical matrix $D(\textbf{q})$ 
The eigenvalue equation of the Hamiltonian (dynamical matrix) is then:
\begin{equation}
%D(\textbf{q}) \textbf{e}_{\textbf{q}n} = \omega^2_{\textbf{q}n} \textbf{e}_{\textbf{q}n},\  \textrm{or}\ 
   \sum_{\beta\kappa'} D_{\kappa\kappa'}^{\alpha\beta}(\textbf{q}) e_{\textbf{q}n}^{\beta\kappa'} = \omega^2_{\textbf{q}n} e_{\textbf{q}n}^{\alpha\kappa},
   \label{dynamic}
\end{equation}
where $n$ is the band index, $\omega_{\textbf{q}n}$ are the phonon frequencies and $e_{\textbf{q}n}^{\alpha\kappa}$ are the phonon eigenvectors in matrix form.

The dynamical matrix $D(\textbf{q})$ is a $3N \times 3N$ matrix, where $3$ comes from the three Cartesian directions and $N$ is the number of atoms in the unit cell. As a result, the eigenvectors $\textbf{e}_{\textbf{q}n}$, which are needed to evaluate the Euler form, are complex column vectors with $3N$ elements
\begin{equation}
    \textbf{e}_{\textbf{q}n}
    =\left(\begin{array}{c}
    e^{x1}_{\textbf{q}n}\\[5pt]
    e^{y1}_{\textbf{q}n}\\[5pt]
    e^{z1}_{\textbf{q}n}\\
        \vdots\\
    e^{xN}_{\textbf{q}n}\\[5pt]
    e^{yN}_{\textbf{q}n}\\[5pt]
    e^{zN}_{\textbf{q}n}
       \end{array} \right),    
\end{equation}
and can be normalized to 1:
\begin{equation}
  \sum_{\alpha\kappa} \big | e_{\textbf{q}n}^{\alpha\kappa} \big | ^ 2 = 1  .
\end{equation} 
We also note that we use the conventional label $\mathbf{q}$ to describe coordinates in the phonon Brillouin zone, rather than the general label $\textbf{k}$ used in Sec.\,\ref{theory}.

The phonon eigenvectors can be used to obtain the associated atomic displacements \cite{Dove1993,Togo2008,Togo2015}
\begin{multline}
[\textbf{u}(l 1), ... , \textbf{u}(l N)]%^{\textrm{T}}
= \\
\bigg[ \frac{A}{\sqrt{m_1}} \textbf{e}^1_{\textbf{q}n} \textrm{e}^{i \textbf{q}\cdot \textbf{r}(l1)}, ... ,
 \frac{A}{\sqrt{m_N}} \textbf{e}^N_{\textbf{q}n} \textrm{e}^{i \textbf{q}\cdot \textbf{r}(lN)}
  \bigg] %^{\textrm{T}},
\end{multline}
where the three-component vectors $\textbf{e}^{\kappa}_{\textbf{q}n}= (e^{x\kappa}_{\textbf{q}n},
e^{y\kappa}_{\textbf{q}n},
e^{z\kappa}_{\textbf{q}n})$, and $A$ is the complex constant \cite{Dove1993,Togo2008,Togo2015}.
%}

%can be normalized to 1

%\textcolor{red}{see mail- A is constant so cannot be a operators. We use eq 5 for Eq 9 jus the plane wave basis. The $U_x$ mean square displacements are quantized there. See anszta 12 in ref 91 that relates to our Eq 7. SO just following it makes it consistent/...}
%\begin{equation}
%  \sum_{\alpha\kappa} \big | e_{\textbf{q}n}^{\alpha\kappa} \big | ^ 2 = 1  ,
%\end{equation}
%and $A$ is the complex constant
%\begin{equation}
%    A = \sqrt{\frac{\hbar}{2N}} \sum_{\textbf{q}n} 
%   \frac{\hat{a}_{\textbf{q}n} \textrm{e}^{-i \omega_{\textbf{q}n}t}
%  +\hat{a}_{-\textbf{q}n}^{\dag} \textrm{e}^{i \omega_{\textbf{q}n}t}
%   }
%   {\sqrt{\omega_{\textbf{q}n}}},
%\end{equation}
%where $\hbar$ is the reduced Planck constant, $\hat{a}_{\textbf{q}n}$ and $\hat{a}_{-\textbf{q}n}^{\dag}$ are the annihilation and creation operators of phonons respectively, and $t$ is the time.

\subsection{Euler class calculations}

After obtaining the phonon dispersion and eigenvectors, we need to find all the phonon band crossing points in gap $\Delta_n$. Then we divide the 2D Brillouin zone into different patches, %that are overlapped with each other, 
and each patch contains either one node or a pair of nodes in $\Delta_n$. The phonon eigenvectors $\textbf{e}_{\textbf{q}n}$ and $\textbf{e}_{\textbf{q}(n+1)}$ are computed on a discretized grid over the patch. We clarify that the patch for nodes in $\Delta_n$ should not overlap with the positions of nodes in the neighboring gaps $\Delta_{n-1}$ and $\Delta_{n+1}$.

The Euler form in Eq.\,(\ref{Euler}) is calculated for all the patches in gap $\Delta_n$ with the phonon eigenvectors $\textbf{e}_{\textbf{q}n}$ and $\textbf{e}_{\textbf{q}(n+1)}$ as input. For each band and $\mathbf{q}$-point, the eigenvector is generally composed of a set of three complex values associated with each atom along the Cartesian axes. In the presence of $C_2T$ symmetry (or $PT$ symmetry) that squares to the identity, there always exists a unitary transformation under which the dynamical matrix becomes real, and the associated eigenvectors are then also real. This unitary transformation is obtained through the Takagi factorization of the matrix representation of the $C_2T$ symmetry, which turns out to be symmetric \cite{Bouhon2020}. Then, isolating the unitary part of the matrix representation of $C_2T$, the Takagi factorization can be readily obtained through singular value decomposition \cite{Chebotarev2014,Chen2021}. We provide an explicit example of this procedure in Sec.~\ref{subsec:takagi} below for aluminium oxide.

Using the real basis we can directly evaluate the expression in Eq~\eqref{Eq:patchEuler}, which is implemented in a publicly available {\sc Mathematica} code \cite{Bzdusek2019}. Alternatively, we note that the patch Euler class of nodes can be calculated by employing Wilson-loop methods~\cite{Bouhon2019, Bouhon2020}. Because of the random sign gauge $+/-$ and the presence of the Dirac string with a gauge transformation, we can smooth the eigenvectors by computing the Berry phase and fixing the position of the Dirac strings. As a result, the eigenvectors $\textbf{e}_{\textbf{q}n}$ and $\textbf{e}_{\textbf{q}(n+1)}$ vary smoothly away from the Dirac strings, with both states flipping their signs simultaneously when crossing a Dirac string. The Berry phase calculations also provide information on the positions and the Berry curvature of the nodes, which helps to verify whether the patch contains the node(s) we are interested in. For details we refer the reader to the Supplementary Material of Ref.~\cite{Bouhon2020}. The gauged eigenvectors can then be used to compute Eq.\,(\ref{Eq:patchEuler}) over the patch $\cal{D}$.

After evaluating the relative stability of each pair of nodes, we can assign the frame charges to all the nodes based on the Euler class of all the patches, as well as the Dirac strings that connect all the linear nodes. We repeat this procedure until a global topological configuration is obtained, as outlined above. 

\section{Case study: phonon braiding in monolayer ${\rm Al_2O_3}$} 
\label{case}

In this section, we exemplify the calculation of the Euler class and associated non-Abelian braiding using first principles methods. To do so, we explore non-Abelian braiding in the phonon spectrum of monolayer ${\rm Al_2O_3}$ as controlled by electrostatic doping.

\subsection{First principles calculations}

Density functional theory (DFT) calculations are performed with the Vienna \textit{ab initio} Simulation Package ({\sc vasp}) \cite{Kresse1996,Kresse1996a}. The generalized gradient approximation (GGA) with the Perdew-Burke-Ernzerhof (PBE) parameterization is used as the exchange-correlation functional \cite{Perdew2008}. A plane-wave basis with a kinetic energy cutoff of $800$\,eV and a $9\times9$ $\mathbf{k}$-mesh are used for monolayer Al$_2$O$_3$. The self-consistent field calculations are stopped when the energy difference between successive steps is below 10$^{-6}$ eV, and the structural relaxation is stopped when forces are below 10$^{-3}$ eV/\AA. A vacuum spacing larger than 20 \AA\ is used to eliminate interactions between adjacent layers. Electrostatic doping is simulated by introducing extra charges with a compensating background. There is no out-of-plane dipole upon electrostatic doping as the extra charge is distributed evenly on the 2D plane (for details, see Supplementary Material). We keep the lattice constants fixed upon doping to mimic the material growth on a substrate, and the ionic positions remain the same upon doping under structural relaxation.

The force constants to determine the phonons are computed using the finite differences method in a $3\times3$ supercell (equivalent to a $3\times3$ phonon $\mathbf{q}$-mesh) with a $3\times3$ electronic $\mathbf{k}$-mesh using {\sc vasp}. The phonon dispersion and phonon eigenvectors are obtained using {\sc phonopy} \cite{Togo2008,Togo2015}. Convergence tests have been performed comparing supercells of sizes between $3\times3$ and $6\times6$ (for details, see Supplementary Material). We also check the convergence of the phonon dispersion in doped Al$_2$O$_3$ with respect to the vacuum spacing, which shows that the phonon frequencies are independent of the vacuum spacing (for details, see Supplementary Material). This is consistent with the fact that no out-of-plane dipole is observed upon doping, and it is therefore sufficient to use the compensating background charge when introducing the extra charges, without the need to include a Coulomb cutoff in the vacuum spacing \cite{Sohier2019}. We focus on hole doping because imaginary phonon modes are observed upon electron doping, indicating that the lattice becomes dynamically unstable in the latter case (for details, see Supplementary Material).
%Hereafter we use the force constants of the $2\times2\times2$ supercell to construct the phonon tight-binding Hamiltonian for simplicity.
The splitting between the longitudinal and transverse optical phonons (LO-TO splitting) is neglected because in 2D materials no LO-TO splitting occurs at $\Gamma$ and only the slope of phonon bands changes \cite{Sohier2017}, which implies that the nodal structure will remain unchanged.

The phonon band crossing points for all the bands are calculated using {\sc WannierTools} \cite{Wu2018}. After obtaining all the nodes in gap $\Delta_n$, we divide the 2D Brillouin zone into different patches, % that overlap with each other, 
and each patch contains either one node or a pair of nodes in $\Delta_n$. Each patch in the 2D Brillouin zone is sampled with a $30\times 30$ $\mathbf{q}$-mesh, and the phonon eigenvectors $\textbf{e}_{\textbf{q}n}$ and $\textbf{e}_{\textbf{q}(n+1)}$ are computed at each of the points sampled, and subsequently rotated to the real basis. The real eigenvectors are then used to calculate the patch Euler class using a modified version of a publicly available {\sc Mathematica} code \cite{Bzdusek2019}, with the calculated real eigenvectors as input.

\subsection{$C_2T$ representation and Takagi factorization} \label{subsec:takagi}

%\textcolor{blue}{Take the underneath and compare to Eq 3 and 7. Then it seems that $\alpha=A/\sqrt{M}$ and you have plus $+{i}$ for the plane wave exponential and $r_{\kappa}=0$. You probably used ref 91. So insert this Eq 10  so that the next basis $C_2T$ transformation is evident and we can submit. }

%{\color{red}
%THIS MUST BE INSERTED AT THE APPROPRIATE PLACE, PLEASE CHECK WHICH EXACT FORM YOU USE IN THE DEFINITION OF $D(\textbf{q})$ AND REMOVE QUESTION MARKS AFTER DECIDING:
%Plane wave ansatz:
%\begin{equation}
%    \textbf{u}(l\kappa) = \alpha \boldsymbol{\varepsilon}_{\kappa}(\textbf{q}) 
 %   {\rm e}^{(\pm ?){\rm i} \textbf{q}\cdot (\textbf{R}_{l}(+\textbf{r}_{\kappa} ?) )} ,
%\end{equation}
%$\alpha$ is a normalization constant, we want $\boldsymbol{\varepsilon}_{\kappa}(\textbf{q}) $ to be a unit vector, and we define $\boldsymbol{\varepsilon}_{\kappa} = (\varepsilon_{x,\kappa},\varepsilon_{y,\kappa},\varepsilon_{z,\kappa})$. 
%}

In this section we detail the steps to transform the phonon eigenvectors to a real basis in the case of monolayer Al$_2$O$_3$. This real representation 
of the eigenvectors is needed to compute the patch Euler class as explained above. 

Before deriving the unitary transformation to the real basis, we first need to derive the generic condition on the dynamical matrix that originates from the $C_2T$ symmetry. For this we use the action of symmetries on the displacements $\textbf{u}(l\kappa)= \big[u_x(l\kappa),u_y(l\kappa),u_z(l\kappa) \big]$ at a given unit cell $l$ and a given atomic site $\kappa$ within the unit cell, i.e.~for $\kappa\in\{1,2,3,4,5\}\equiv\{{\rm Al}_1,{\rm Al}_2,{\rm O}_1,{\rm O}_2,{\rm O}_3\}$. It will be convenient to use the ket-form of the displacement vectors
\begin{equation}
\textbf{u}(l\kappa)^t = \big[ u_x(l\kappa)~
u_y(l\kappa)~u_z(l\kappa) \big].
\end{equation}
Since the system here has both $C_2$ (rotation by $\pi$ around the $\hat{z}$ axis) and $T$ (time reversal) symmetries, we consider the action of each symmetry separately. The action of $C_2$ on the displacements gives
\begin{equation}
\label{eq_sym_1}
    ^{C_2}\textbf{u}(l\kappa)^t = \textbf{u}(C_2[l\kappa])^t \cdot\Gamma^{({\rm vec})}(C_2),
\end{equation}
where $\Gamma^{({\rm vec})}(C_2)$ is the vector representation of the point group of the system for $C_2$, i.e.
\begin{equation}
\label{eq_sym_1a}
    \Gamma^{({\rm vec})}(C_{2z}) = {\rm diag}(-1,-1,1),
\end{equation}
while $C_2[l\kappa]$ stands for
\begin{equation}
\label{eq_sym_2}
      C_2 \textbf{r}(l\kappa) = \textbf{r}(l'\kappa') = \textbf{r}(l\kappa') + \boldsymbol{\Delta}_{ll'},
\end{equation}
where $\boldsymbol{\Delta}_{ll'} = \textbf{R}_{l'}-\textbf{R}_{l}$ is a Bravais vector and $\textbf{r}(l\kappa')$ is the position of the $\kappa'$-th atomic site within the $l$-th unit cell determined through the permutation 
\begin{equation}
\label{eq_sym_3}
\begin{aligned}
    C_2({\rm Al}_1~{\rm Al}_2~{\rm O}_1~{\rm O}_2~{\rm O}_3) &=
    ({\rm Al}_1~{\rm Al}_2~{\rm O}_1~{\rm O}_2~{\rm O}_3) \cdot U_{{\rm perm}}\\
    &= ({\rm Al}_2~{\rm Al}_1~{\rm O}_1~{\rm O}_2~{\rm O}_3),
\end{aligned}
\end{equation}
where $U_{{\rm perm}} = \sigma_x \oplus \mathbb{1}_{3}$. The action of $T$ simply gives
\begin{equation}
\label{eq_sym_4}
      ^{T}\textbf{u}(l\kappa)^t = \mathcal{K}\textbf{u}(l\kappa)^t = \textbf{u}(l\kappa)^t \mathcal{K},
\end{equation}
where $\mathcal{K}$ is the complex conjugation, and the last equality follows from the assumption that the displacements are real. Let us now rewrite the displacement in the Bloch form, 
\begin{equation}
    \textbf{u}(l\kappa) = \sum_{\textbf{q}\in {\rm BZ}}
    {\rm e}^{{\rm i} \textbf{q}\cdot \textbf{r}(l\kappa)} \boldsymbol{\varepsilon}_{\textbf{q}}(\kappa),
\end{equation}
(BZ is the Brillouin zone), through which the potential energy takes the form,
\begin{multline}
    \sum\limits_{ll'}
    \textbf{u}(l\kappa)^t \cdot \boldsymbol{\Phi}(l\kappa,l'\kappa') \cdot \textbf{u}(l\kappa) \propto \\ \sum_{\textbf{q}\in {\rm BZ}}
    \boldsymbol{\varepsilon}_{\textbf{q}}(\kappa)^{\dagger} \cdot D_{\kappa\kappa'}(\textbf{q}) \cdot \boldsymbol{\varepsilon}_{\textbf{q}}(\kappa'),
\end{multline}
where we have used $\boldsymbol{\varepsilon}_{\textbf{q}}(\kappa)^* = \boldsymbol{\varepsilon}_{-\textbf{q}}(\kappa)$ that follows from the reality of $\textbf{u}(l\kappa)$. Then, we get the $C_2T$ symmetry constraint on the dynamical matrix from the symmetry action on the Bloch components $\boldsymbol{\varepsilon}_{\textbf{q}}(\kappa)$. 

Ordering the Bloch components into a $(1\times 15)$ complex vector, i.e.
\begin{equation}
    \vert \boldsymbol{\varepsilon}_{\textbf{q}} \rangle = \big[ \varepsilon_{\textbf{q},x}(1) ~\varepsilon_{\textbf{q},y}(1)~
    \varepsilon_{\textbf{q},z}(1) \ ...\  \varepsilon_{\textbf{q},x}(5) ~\varepsilon_{\textbf{q},y}(5)~
    \varepsilon_{\textbf{q},z}(5) \big] ^*,
\end{equation}
the $\textbf{q}$-component of the potential energy takes the form
\begin{equation}
    \vert \boldsymbol{\varepsilon}_{\textbf{q}} \rangle \cdot D(\textbf{q}) \cdot \langle \boldsymbol{\varepsilon}_{\textbf{q}} \vert, 
\end{equation}
and from Eq.(\ref{eq_sym_1})-(\ref{eq_sym_4}) we get the $C_2T$ symmetry action  
\begin{equation}
    \begin{aligned}
        ^{C_2T}\vert\boldsymbol{\varepsilon}_{\textbf{q}}\rangle = \vert\boldsymbol{\varepsilon}_{-C_2\textbf{q}}\rangle  \cdot U_{C_2T} \mathcal{K} = \vert\boldsymbol{\varepsilon}_{\textbf{q}}\rangle  \cdot U_{C_2T} \mathcal{K},
    \end{aligned}
\end{equation}
with 
\begin{equation}
   U_{C_2T} = U_{{\rm perm}} \otimes 
   \Gamma^{({\rm vec})}(C_2).
\end{equation}
The invariance of the potential energy under $C_2T$ symmetry then gives the following constraint on the dynamical matrix
\begin{eqnarray}
    ^{C_2T}(\Phi-\Phi_0) &=& (\Phi-\Phi_0) ,\nonumber\\
    \Leftrightarrow \quad
    U_{C_2T}\cdot \big[D(\textbf{q}) \big] ^* \cdot U_{C_2T}^{\dagger}
    &=& 
     D(\textbf{q}).
\label{eq_sym}
\end{eqnarray}

First, we verify that the double action of $C_2T$ gives an identity, i.e.
\begin{equation}
    \langle \boldsymbol{\varepsilon}_{\textbf{q}}\vert ^{[C_2T]^2}\vert\boldsymbol{\varepsilon}_{\textbf{q}}\rangle = U_{C_2T}\cdot \left(U_{C_2T}\right)^* = \mathbb{1}_{15}.
    %\textbf{e}_{\textbf{q}n}\cdot  \left(^{[C_2T]^2}(\textbf{e}_{\textbf{q}n})^t\right) = U_{C_2T}\cdot (U_{C_2T})^* =  \mathbb{1}_{15}
\end{equation}
Then, by unitarity $(U_{C_2T})^* = \big[\left(U_{C_2T}\right)^t\big]^{-1}$, and we readily obtain that the matrix is symmetric, i.e.~$U_{C_2T} = (U_{C_2T})^t$. We can therefore perform a Takagi factorization, given by $U_{C_2T} = U_{{\rm tf}}\cdot \Lambda \cdot (U_{{\rm tf}})^t$ where $\Lambda$ is diagonal, from which we define the unitary matrix $W=\sqrt{\Lambda^*} \cdot (U_{{\rm tf}})^{\dagger}$. We describe below how to derive the unitary matrices $\Lambda$ and $U_{{\rm tf}}$. 

We now define the rotated basis $\widetilde{\boldsymbol{\varepsilon}}$ through
\begin{equation}
    \vert\boldsymbol{\varepsilon}_{\textbf{q}}\rangle =
    \vert\widetilde{\boldsymbol{\varepsilon}}_{\textbf{q}}\rangle
    \cdot W,
\end{equation}
in which the dynamical matrix is real. Indeed, the representation of $C_2T$ in the new basis reads 
\begin{equation}
    ^{C_2T}\vert\widetilde{\boldsymbol{\varepsilon}}_{\textbf{q}}\rangle = \vert\widetilde{\boldsymbol{\varepsilon}}_{\textbf{q}}\rangle \cdot \left( 
        W U_{C_2T} W^t
    \right) \mathcal{K} = \vert\widetilde{\boldsymbol{\varepsilon}}_{\textbf{q}}\rangle \mathcal{K},
\end{equation}
i.e.~the unitary part of $C_2T$ is now a unit matrix. Rotating the dynamical matrix in the new basis, i.e.~
\begin{equation}
    W \cdot D(\textbf{q})\cdot W^t =\widetilde{D}(\textbf{q})  ,    
\end{equation}
the $C_2T$ symmetry constraint Eq.\,(\ref{eq_sym}) becomes
\begin{equation}
     \big[\widetilde{D}(\textbf{q})\big]^*  = \widetilde{D}(\textbf{q}).
\end{equation}
We conclude that the eigenvectors of $\widetilde{D}(\textbf{q})$ must be real (and symmetric). 

We end with the derivation of $W$. Because $U_{C_2T}$ is unitary on top of being symmetric, the Takagi factorization is readily given through a singular value decomposition \cite{Chebotarev2014}, i.e.~$U_{C_2T} = U_{{\rm svd}} \cdot \Lambda \cdot V_{{\rm svd}} $ with $\Lambda = \mathbb{1}$, from which we get $U_{{\rm tf}} = U_{{\rm svd}} \cdot \sqrt{(U_{{\rm svd}})^{\dagger} \cdot (V_{{\rm svd}})^*}$. The unitary matrix that rotates to the new basis is finally given by $W = (U_{{\rm tf}})^{\dagger}$. 

We note that the above derivation is completely general, with only $U_{C_2T}$ being system dependent (for another example of this procedure applied to an electronic band structure problem, see Ref.~\cite{Chen2021}). 

\subsection{Crystal structure and phonon dispersion}

\begin{figure}
\centering
\includegraphics[width=0.8\linewidth]{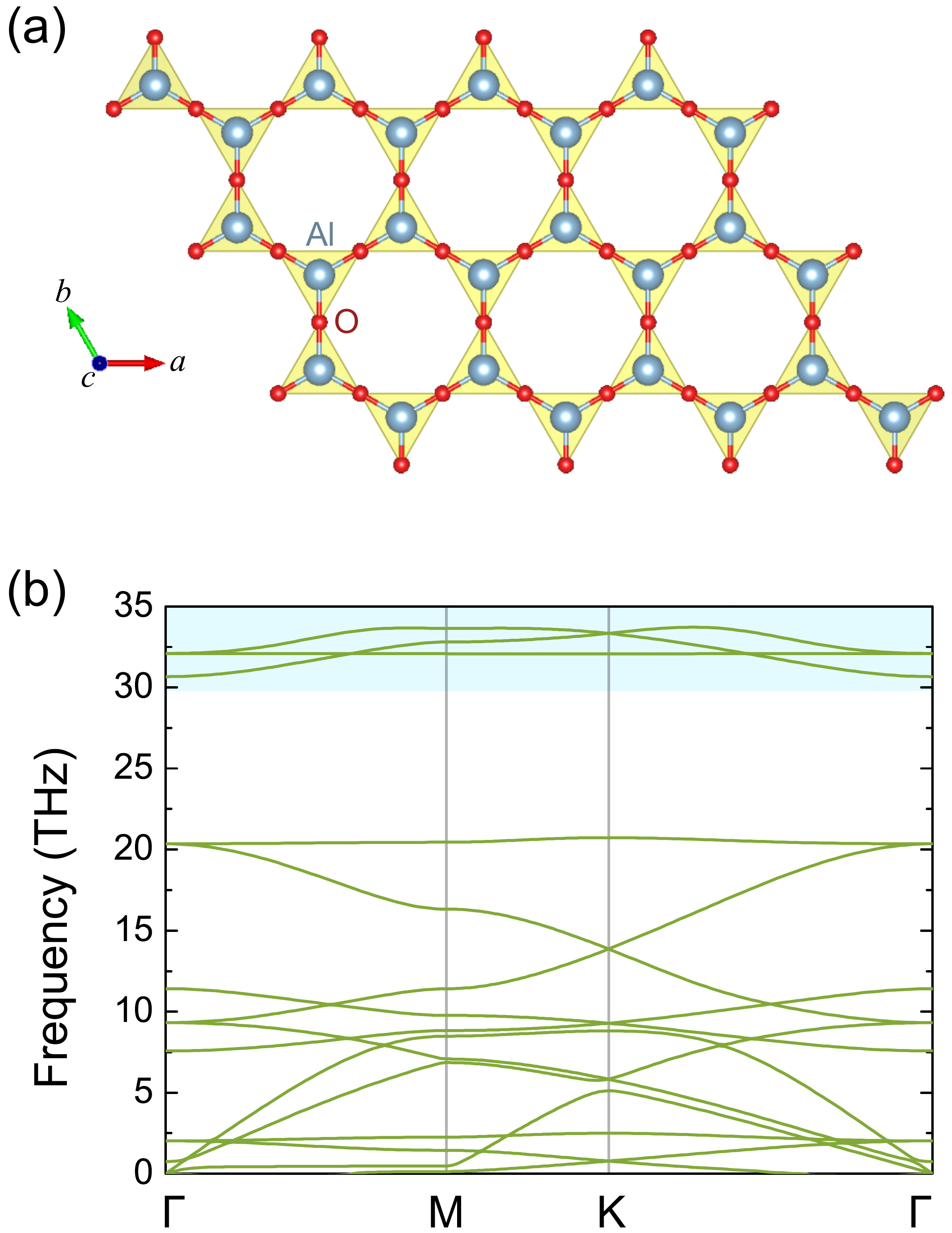}
\caption{(a) Crystal structure and (b) phonon dispersion of monolayer Kagome Al$_2$O$_3$.}
\label{crystal} 
\end{figure}

Monolayer Al$_2$O$_3$ is predicted to crystallize in a honeycomb lattice \cite{Song2016a}. The aluminum and oxygen atoms are in the same plane, with the oxygen atoms forming a Kagome lattice in 2D, as shown in Fig.~\ref{crystal}(a). The calculated lattice constant of 5.842 \AA\ agrees well with previous calculations \cite{Song2016a}. Monolayer Al$_2$O$_3$ belongs to the $P6/mmm$ space group (No.\,191), which has $C_2$ rotation symmetry. In addition, in phonons the time reversal symmetry $T$ is automatically satisfied. With $C_2T$ symmetry, phonons in monolayer Al$_2$O$_3$ can be described by a real Hamiltonian (dynamical matrix), and consequently we can assign non-Abelian frame charges to different nodes in any three-band subsystem in the entire phonon spectrum.

Figure~\ref{crystal}(b) shows the calculated phonon dispersion. No imaginary phonon modes are observed, indicating the dynamical stability of monolayer Al$_2$O$_3$. There are 5 atoms in the unit cell, leading to 15 phonon branches. We focus on the top three bands, i.e.~bands 13$-$15 marked in the blue area between $30$ and $35$ THz in Figure~\ref{crystal}(b), because they are isolated from other phonon bands and are more sensitive to electrostatic doping (for details, see Supplementary Material), thus providing an ideal platform to explore multi-gap topology and non-Abelian braiding.

\subsection{Band inversion upon electrostatic doping}

\begin{figure}
\centering
\includegraphics[width=0.93\linewidth]{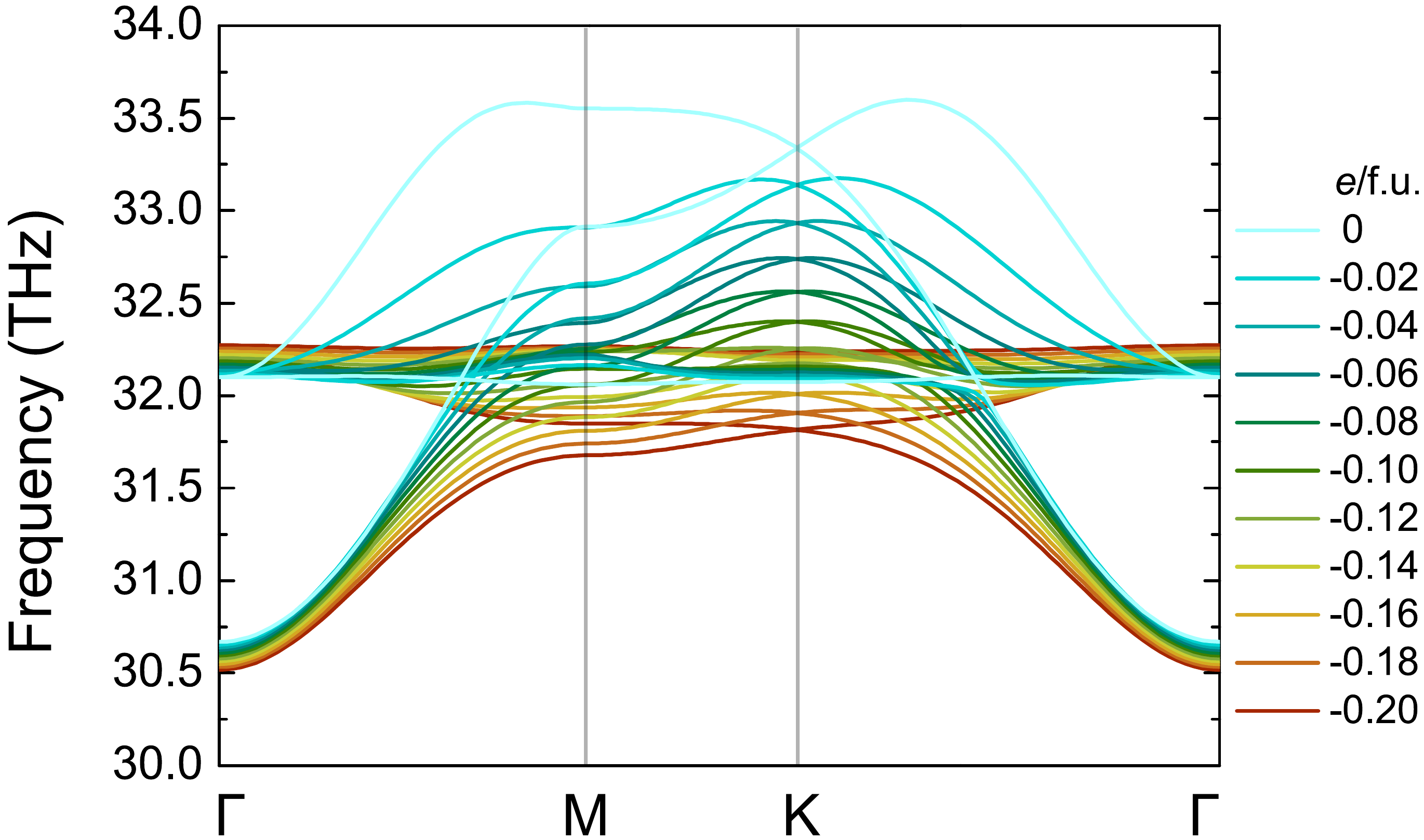}
\caption{Evolution of phonon bands $13-15$ upon electrostatic doping.}
\label{doping} 
\end{figure}

Bulk Al$_2$O$_3$ is a well-known dielectric material used in electronic devices. Therefore, electrostatic doping of monolayer Al$_2$O$_3$ by gate voltage can be easily incorporated into the existing microelectronics industry. We simulate the phonon spectra of Al$_2$O$_3$ upon electrostatic doping. As shown in Fig.~\ref{doping}, the phonon frequencies of bands 13$-$15 at the $\Gamma$ point have only slight changes upon doping. On the other hand, the highest phonon bands at the K point with double degeneracy move to much lower frequency with increasing doping concentration, whereas the frequency of the non-degenerate single band at K remains nearly the same. Therefore, the band order at K is inverted at $-0.14$ $e$/f.u., with the double degenerate band becoming lower than the single band.

Such phonon band inversion redistributes the band nodes in two neighboring gaps $\Delta_{13}$ and $\Delta_{14}$, and consequently induces conversions of the frame charges. The transfer of frame charges between different gaps is accompanied by the non-Abelian braiding of the nodes. It is however important to note that the crystalline symmetries of the system constrain the movements of the nodes over the Brillouin zone with, as a consequence, the collapse of the braid trajectories onto the high-symmetry points $\Gamma$ and K. In the next subsections we first investigate the topological configurations at different doping concentrations individually, and then obtain the complete picture of the braiding processes upon electrostatic doping.

%{\color{red} Shall we show all the phonon spectra and the corresponding topological configurations at different doping ratios first? Or we show the configuration at $-0.08$ e/f.u. first (Fig.~8), then at $-0.10$ e/f.u. etc, and show the braiding processes in the last?}

\subsection{Frame charges of undoped Al$_2$O$_3$}

\begin{figure}
\centering
\includegraphics[width=0.8\linewidth]{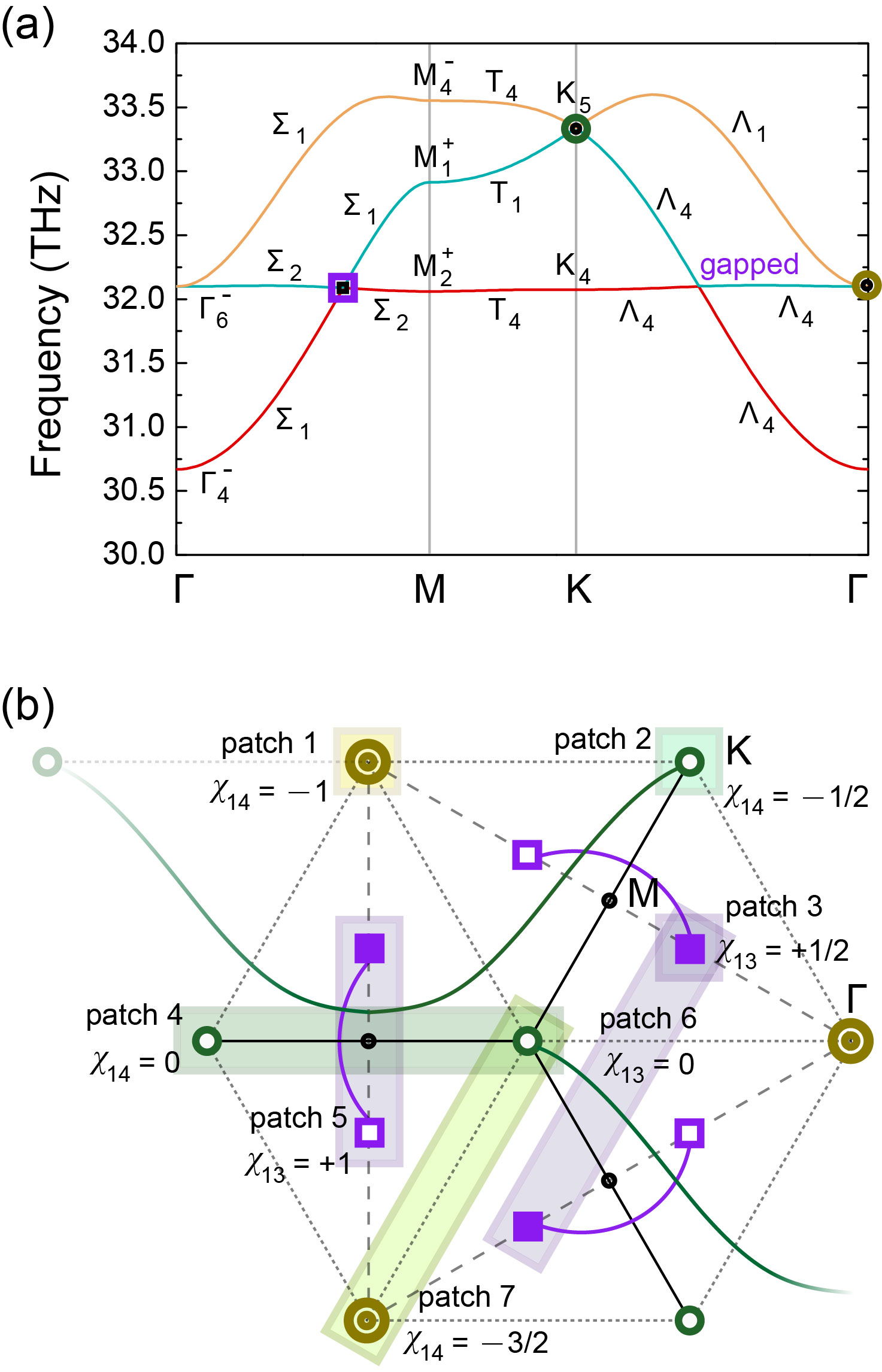}
\caption{(a) Phonon band crossing points and (b) patch Euler class of undoped Al$_2$O$_3$. We use squares (circles) to represent the nodes formed by the lower (upper) two bands, open (closed) symbols to represent the nodes with negative (positive) frame charges, and one symbol (two concentric symbols) to represent the linear (quadratic) node.}
\label{patch-0} 
\end{figure}

We first investigate the phonon band crossing points formed by band 13, 14 and 15 of undoped Al$_2$O$_3$. To be consistent with the notation introduced in the theoretical background section (Section~\ref{theory}), we use squares to represent single nodes in gap $\Delta_{13}$ formed by bands 13 and 14, and circles for single nodes in gap $\Delta_{14}$ formed by bands 14 and 15. As shown in Fig.~\ref{patch-0}(a), a band crossing point is formed when the two crossing bands belong to different irreducible representations (irreps), whereas two bands with the same irrep remain gapped. To be specific, the violet square node along $\Gamma$-M formed by bands 13 and 14 [red and blue lines in Fig.~\ref{patch-0}(a)] is not gapped as the two bands belong to irreps $\Sigma_1$ and $\Sigma_2$, but these two bands have an avoided crossing along K-$\Gamma$ because in that case they belong to the same irrep $\Lambda_4$. In addition, there are two circle nodes at $\Gamma$ and K formed by bands 14 and 15 [blue and yellow lines in Fig.~\ref{patch-0}(a)] with 2D irreps $\Gamma_6^-$ and K$_5$ respectively.

%Such charges $q$ are the frame rotations of real phonon eigenvectors, and can be characterized by the non-Abelian quaternion group \cite{Wu2019a}, i.e.~$q\in \{1,i,j,k\}$ with $i^2=j^2=k^2=-1$, $ij=k$, $jk=i$ and $ki=j$. 

%We can manipulate the non-Abelian charge by momentum space braiding of band nodes in two adjacent gaps, and obtain topological configurations where the pairs of band nodes in the same gap have the same charge. As a result, the band nodes cannot annihilate with each other unless by the inverse braiding process. We can use a new topological invariant to describe whether the band nodes can be annihilated or not. Such topological invariants are called Euler class, and can be computed over patches that contain different pairs of nodes. We first need to compute the phonon eigenvectors $\vert e_{n} \rangle$ and $\vert e_{n+1} \rangle$ for band $n$ and $n+1$ in a patch containing a pair of nodes. The Euler curvature of that patch can be defined as \cite{Zhao2017b,Ahn2018,Ahn2019a,Bouhon2020}
%\begin{equation}
%    {\rm Eu} = (\langle \partial_{k_a} e_{n} \vert \partial_{k_b} e_{n+1} \rangle - \langle \partial_{k_b} e_{n} \vert \partial_{k_a} e_{n+1} \rangle ) dk_a\wedge dk_b,
%\end{equation}
%where a dense $30\times30$ $\mathbf{k}$-mesh is used.  The patch Euler class $\chi_n$ for the pair of nodes formed by band $n$ and $n+1$ can then be calculated as
%\begin{equation}
%\chi_n = \frac{1}{2\pi}  %\int_{{\rm BZ}} {\rm Eu}.
%\end{equation}

Figure~\ref{patch-0}(b) shows the position of all the nodes. We first compute the Euler class for single nodes, as indicated by patches 1$-$3 in Fig.~\ref{patch-0}(b). In patch 1, the dark yellow circle at the $\Gamma$ point has an Euler class of $-1$, indicating a quadratic node. This agrees well with the quadratic dispersion near $\Gamma$. Consistent with Section~\ref{theory}, we label the quadratic node by a small circle inside a large circle because it can be viewed as two linear nodes merged together. On the other hand, the Euler class for the dark green circle in $\Delta_{14}$ at the K point (patch 2) and the violet square in $\Delta_{13}$ along the $\Gamma$-M high-symmetry line (patch 3) is $\pm 1/2$, indicating two linear nodes. Note that for a single node the sign of the Euler class has no physical meaning due to the $+/-$ sign freedom, but the relative signs of two nodes in the same gap provide information on their stability. To be consistent with Section~\ref{theory}, we use open (closed) symbols to represent the nodes with negative (positive) frame charges, as shown in Fig.~\ref{patch-0}(b).

We next compute the Euler class for all the patches containing pairs of \textit{linear nodes}. For the neighboring dark green circles in patch 4, $\chi_{14} = 0$, indicating that these two nodes can either carry opposite frame charges or have the same frame charge with a nearby Dirac string in $\Delta_{13}$. For convenience, we connect the neighboring pair of violet nodes in $\Delta_{13}$ in patch 5 with a violet Dirac string, and assign the same frame charge to the dark green nodes in patch 4. We also connect the pair of dark green nodes with a dark green Dirac string. 

We then calculate the Euler class for all the patches containing all neighboring pairs of the violet square nodes in $\Delta_{13}$. For patch 5, $\chi_{13} = 1$, indicating that the two nearest nodes along $\Gamma$-M can either carry the same frame charge or have opposite frame charges with a nearby Dirac string in $\Delta_{14}$. Because of the presence of a dark green Dirac string in their neighboring gap $\Delta_{14}$ that crosses patch 5, we can assign the opposite frame charges to the violet nodes in $\Delta_{13}$ in patch 5. %and assign the same frame charge to other nearest pairs of violet nodes along $\Gamma$-M. 
In patch 6, $\chi_{13} = 0$, and we can assign the same frame charges to the corresponding violet nodes as there is a Dirac string in $\Delta_{14}$ crossing patch 6.
%However, the two neighboring violet square nodes in patch 5 have to carry the same positive charge because of the presence of a Dirac string in their neighboring gap $\Delta_{14}$.
%To make the calculated $\chi_{13} = 0$ for patch 2 consistent, we add a Dirac string connecting the two dark green circle nodes in $\Delta_{14}$ at $\Gamma$. We
We then assign the Dirac strings for the neighboring violet squares, and obtain the complete global topological configuration shown in Fig.~\ref{patch-0}(b).

Finally, we check the consistency of the global topological configuration by computing the Euler class for patch 7. The calculated $\chi_{14} = -3/2$ is consistent with the presence of a quadratic node with $\chi_{14} = -1$ at $\Gamma$ and a linear node with $\chi_{14} = -1/2$. %as well as a Dirac string in $\Delta_{13}$ that flips the sign of the linear node.

\subsection{Frame charges of $-0.08$ $e$/f.u. doped Al$_2$O$_3$}

\begin{figure}
\centering
\includegraphics[width=0.8\linewidth]{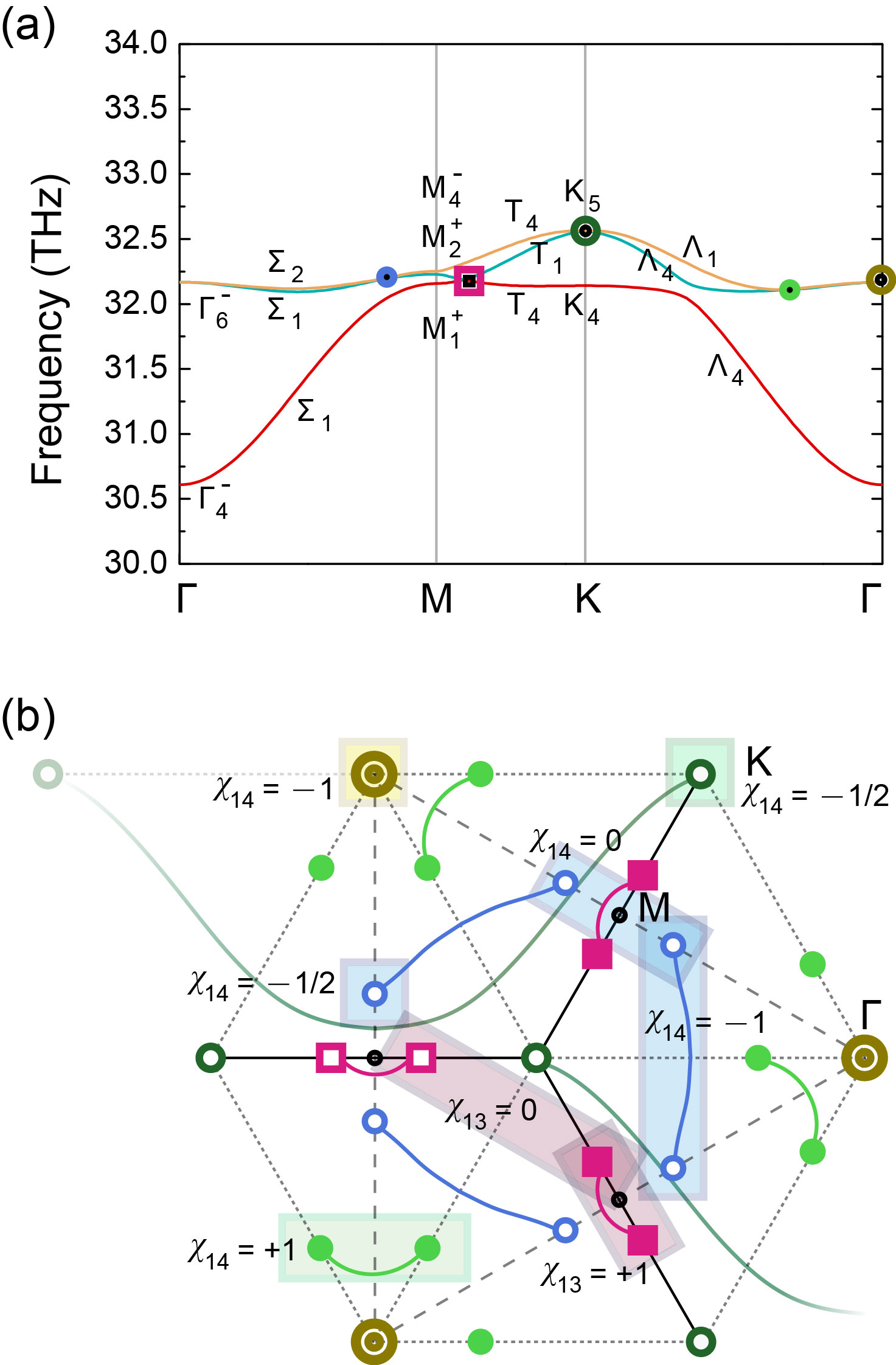}
\caption{(a) Phonon band crossing points and (b) patch Euler class of $-0.08$ $e$/f.u. doped Al$_2$O$_3$.}
\label{patch-8} 
\end{figure}

At $-0.08$ $e$/f.u., several band inversions take place around the $\Gamma$ and M high-symmetry points, as shown in Fig.~\ref{patch-8}(a). Along $\Gamma$-M the top two bands with irreps $\Sigma_1$ and $\Sigma_2$ start to be inverted, and the top two bands along K-$\Gamma$ with irreps $\Lambda_1$ and $\Lambda_4$ are inverted as well. Because these two bands along both $\Gamma$-M and K-$\Gamma$ have different irreps, there are two new circle nodes in $\Delta_{14}$ formed along these two high-symmetry lines, and we label the nodes along $\Gamma$-M in blue and those along K-$\Gamma$ in green. In addition, at the M point, bands 13 and 14 are also inverted, as the band with irrep M$_1^+$ now becomes lower than that with irrep M$_2^+$. As a result, the nodes in $\Delta_{13}$ transfer from $\Gamma$-M to M-K. These band inversions significantly change the number and positions of the band nodes in gaps $\Delta_{13}$ and $\Delta_{14}$, and the distribution of the topological frame charges is completely different from that in undoped Al$_2$O$_3$.

We start the Euler class calculations from the patches containing single nodes. The Euler class of the quadratic node at $\Gamma$ and the linear node at K remains the same. All other nodes, created by the band inversions, are linear nodes as their patch Euler class is $\pm 1/2$.

For the pink square nodes in $\Delta_{13}$ along M-K, the Euler class for the two pink patches, containing the first and second nearest neighbors in Fig.~\ref{patch-8}(b), is 1 and 0 respectively. %From the undoped case in Fig.~\ref{patch-0}(b), we already know that there is a Dirac string connecting the green circles in $\Delta_{14}$. 
Because there is no Dirac string in $\Delta_{14}$ crossing these two patches, we can assign the same frame charge for all the nearest pairs of pink square nodes, while keeping the second nearest pairs either with the opposite frame charges or with the same frame charge and a nearby Dirac string in $\Delta_{14}$.

We then calculate the Euler class for pairs of the neighboring green circles in $\Delta_{14}$, and obtain $\chi_{14} = 1$ for all the patches. Because the green circle nodes are far away from the square nodes in $\Delta_{13}$ (and their Dirac strings), we can safely assign the same frame charge for all the green circle nodes.

For the blue circle nodes along $\Gamma$-M, the patch Euler class for the nearest pair around M is 0, while the second nearest neighbor of nodes has a patch Euler class of $\chi_{14} = -1$. Thus we can assign all the blue circles with the same negative frame charge, with a Dirac string in $\Delta_{13}$ connecting the two pink square nodes around M. We can also connect the second nearest neighbors of the blue circle nodes in pairs with the Dirac strings, without influencing the topological configurations in $\Delta_{13}$.

 The consistent global topological configuration is summarized in Fig.~\ref{patch-8}(b), which is also consistent with the conversion of frame charges from undoped Al$_2$O$_3$ to $-0.08$ $e$/f.u. doped Al$_2$O$_3$ (as discussed later).

%{\color{red} How do the topological configurations evolve from undoped to $$-0.08$$ $e$/f.u. doped Al$_2$O$_3$???}

\subsection{Frame charges of $-0.10$ $e$/f.u. doped Al$_2$O$_3$}

\begin{figure}
\centering
\includegraphics[width=0.8\linewidth]{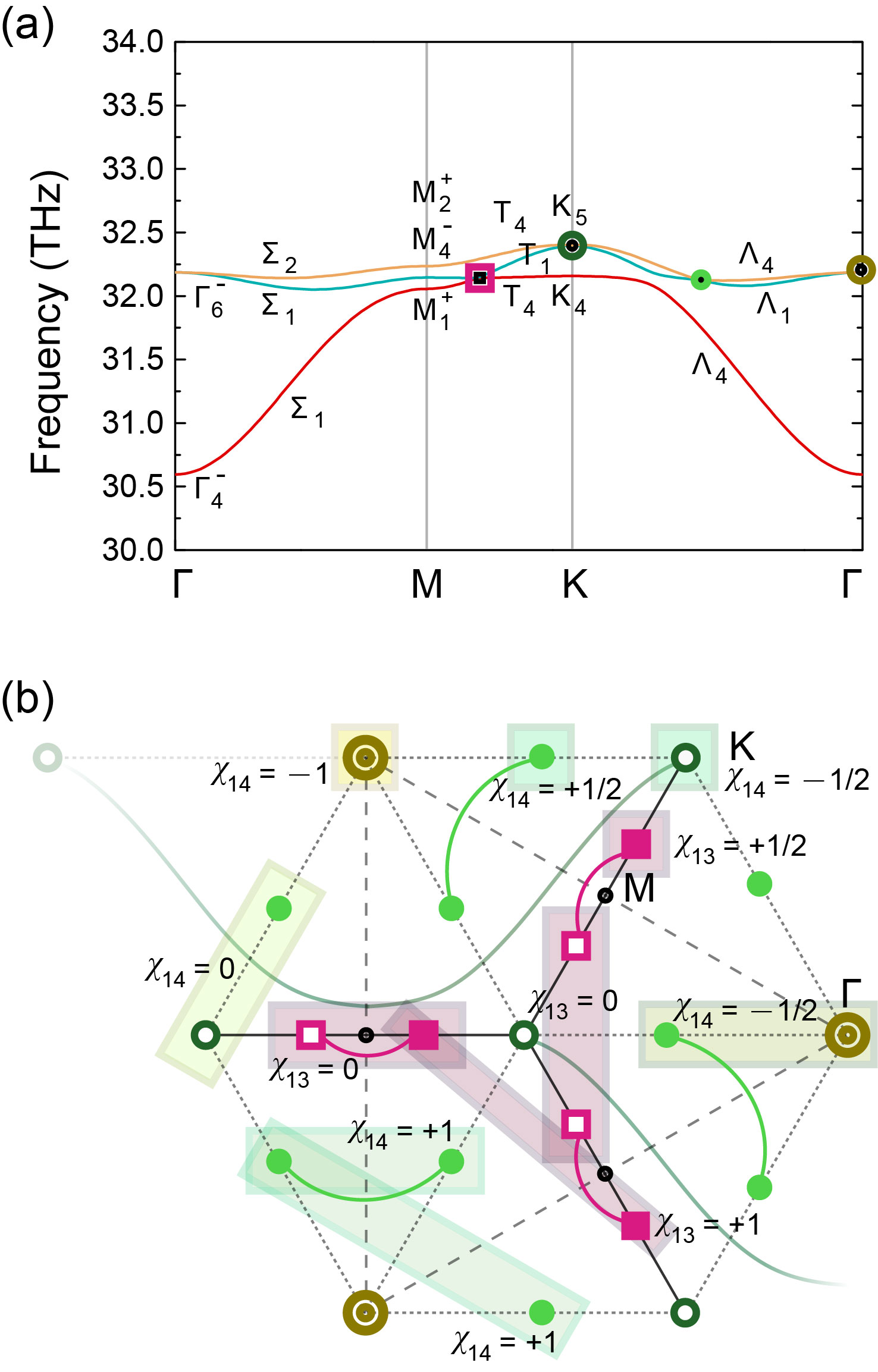}
\caption{(a) Phonon band crossing points and (b) patch Euler class of $-0.10$ $e$/f.u. doped Al$_2$O$_3$.}
\label{patch-10} 
\end{figure}

At $-0.10$ $e$/f.u., the top two bands are inverted at M, and the M$_2^+$ band becomes higher than the M$_4^-$ band [Fig.~\ref{patch-10}(a)]. Consequently the top two bands along $\Gamma$-M are fully inverted as well, as the $\Sigma_2$ band is the highest all along the $\Gamma$-M high-symmetry line. As a result, the blue circle nodes in $\Delta_{14}$ along $\Gamma$-M disappear.

As the inversion of bands 14 and 15 occurs at M, the two nearest blue nodes in $\Delta_{14}$ meet each other at M. Before they meet at M, one of them must cross the pink Dirac string in $\Delta_{13}$, which flips the sign of its frame charge. Therefore, the two blue nodes, now with the opposite frame charge, can annihilate when brought together at M.  The Dirac strings of three pairs of the blue nodes now merge into a closed loop connecting the three neighboring M points, encircling the K point in the middle. By shrinking the closed Dirac string towards K, we can make it disappear, and this also flips the sign of the frame charges of the three pink nodes inside the loop. Consequently, the two nearest pink square nodes along M-K now carry opposite frame charges.

To verify this we compute the patch Euler class in Fig.~\ref{patch-10}(b). The calculated Euler class for the pink nodes is consistent with our deduction that the two nearest pink nodes have opposite frame charges. In addition, as the braiding only takes place around M, the topological frame charges remain unchanged for other nodes away from M. This is also confirmed by our Euler class calculations.

\subsection{Frame charges of -0.20 $e$/f.u. doped Al$_2$O$_3$}

\begin{figure}
\centering
\includegraphics[width=0.8\linewidth]{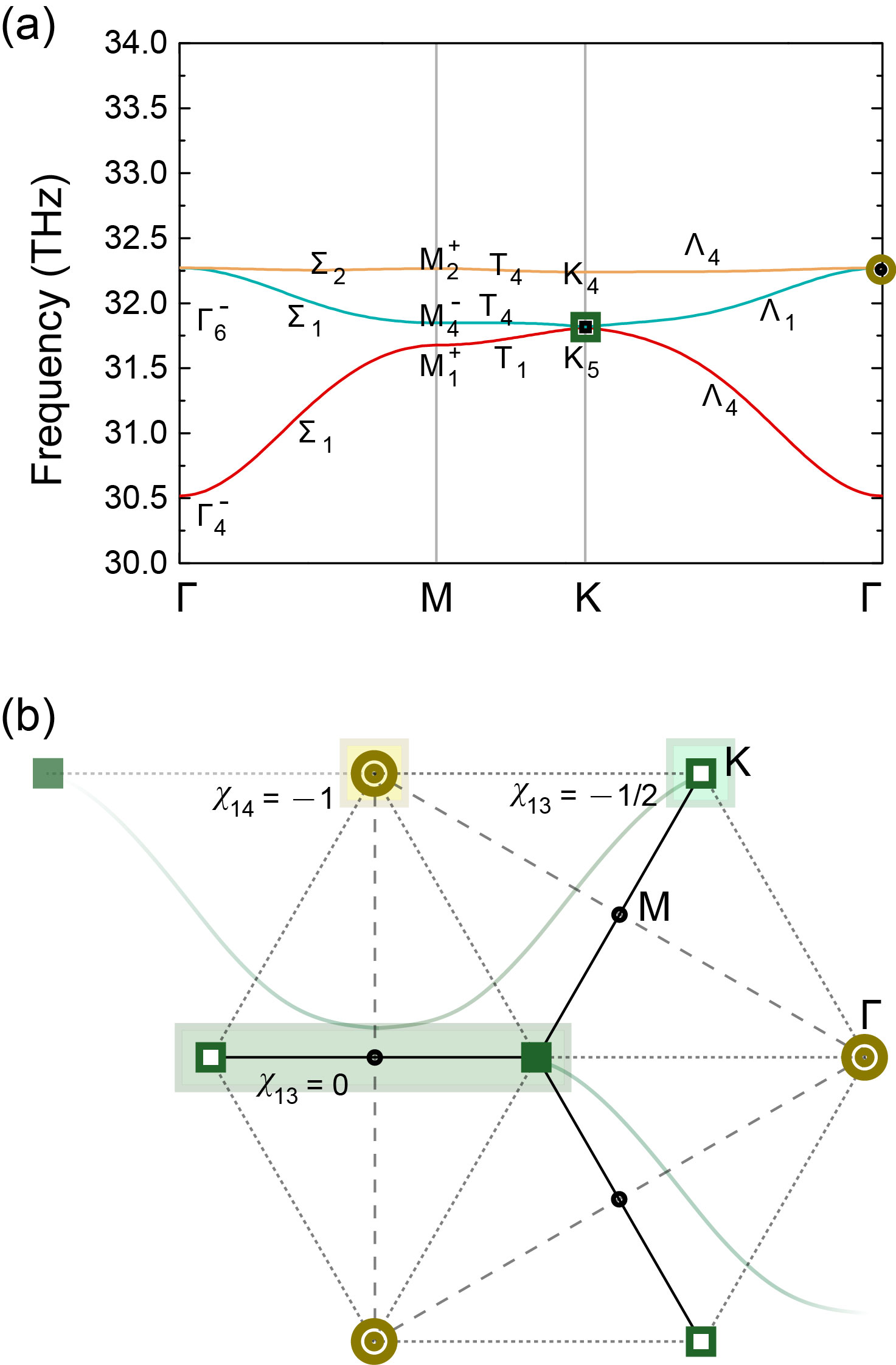}
\caption{(a) Phonon band crossing points and (b) patch Euler class of $-0.20$ $e$/f.u. doped Al$_2$O$_3$.}
\label{patch-20} 
\end{figure}

At $-0.14$ $e$/f.u., the doubly degenerate band with K$_5$ irrep becomes lower than the K$_4$ band, and upon further doping the topological configurations remain the same as no additional band inversion occurs. We show the phonon dispersion and the corresponding global topological configuration at $-0.20$ $e$/f.u. in Fig.~\ref{patch-20} because the bands are well separated from each other at this doping density so we can label the irreps more clearly.

%{\color{red} How do the topological configurations in $-0.12$ $e$/f.u. convert to that in -0.20 $e$/f.u.???}

We compute the Euler class for all the single nodes first, and the calculated $\chi_{14} = -1$ at $\Gamma$ and $\chi_{13} = -1/2$ at K indicate a robust quadratic node in $\Delta_{14}$ at $\Gamma$ and a linear node in $\Delta_{13}$ at K. We also compute the Euler class for the patch containing two nearest K points, and obtain $\chi_{13} = 0$. Therefore we can assign opposite frame charges to the K points, and connect each pair of them with a Dirac string.

\subsection{Complete picture of braiding upon doping}

\begin{figure*}
\centering
\includegraphics[width=\linewidth]{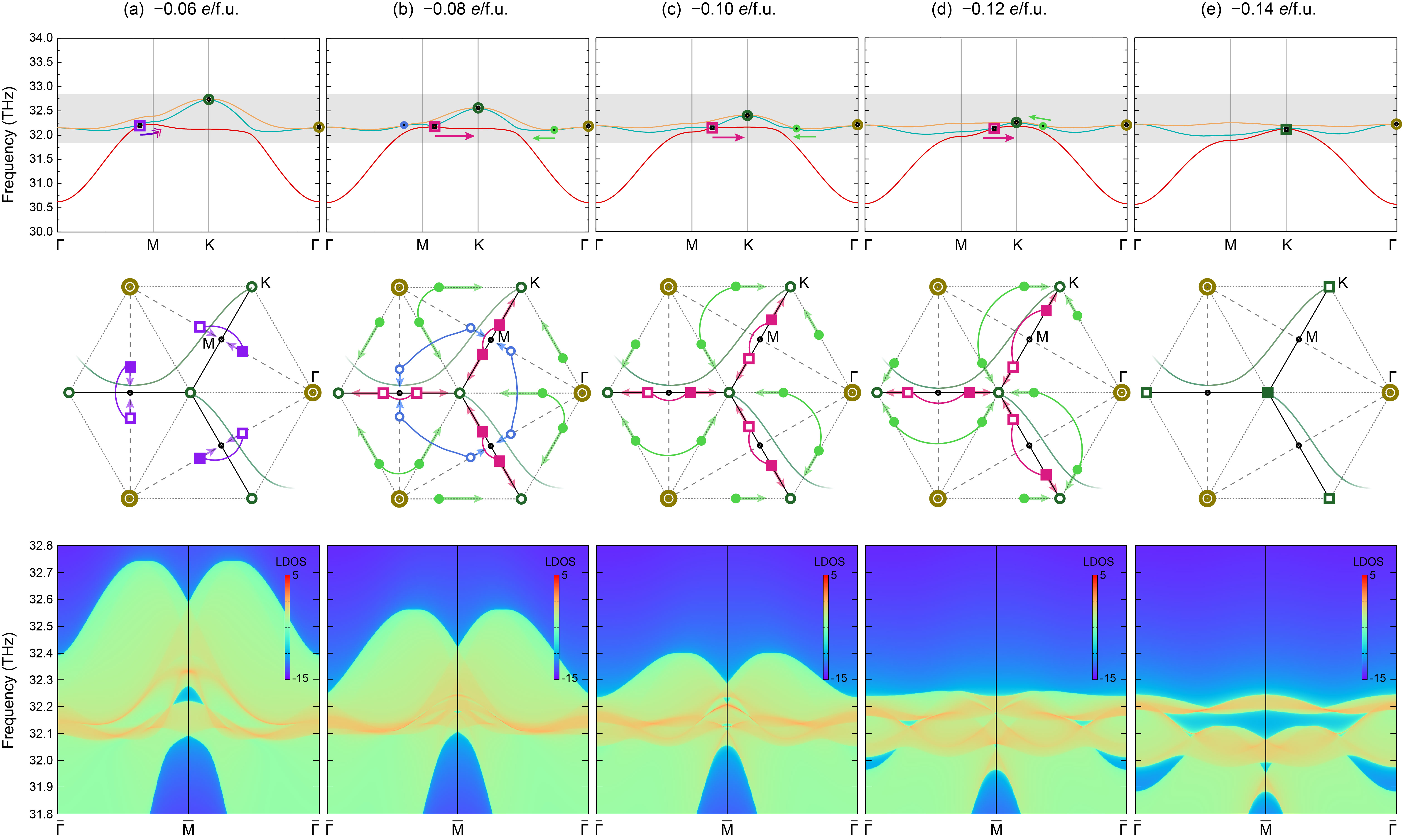}
\caption{Phonon spectra (top panel) and topological configurations (middle panel) of monolayer Al$_2$O$_3$ upon electrostatic doping at (a) $-0.06$ $e$/f.u., (b) $-0.08$ $e$/f.u., (c) $-0.10$ $e$/f.u., (d) $-0.12$ $e$/f.u., and (e) $-0.14$ $e$/f.u. For the topological configurations in the middle panel, we use squares (circles) to represent the nodes formed by the lower (upper) two bands, open (closed) symbols to represent the nodes with negative (positive) frame charges, and one symbol (two concentric symbols) to represent the linear (quadratic) node. The edge states along the (100) direction are also shown in the bottom panel, corresponding to the grey area of the phonon spectra in the top panel.}
\label{allphonon} 
\end{figure*}

The complete picture, summarized in Fig.~\ref{allphonon}, provides a detailed description of the conversion of non-Abelian frame charge upon electrostatic doping from $-0.06$ to $-0.14$ $e$/f.u.
%{\color{red} Shall we remove Fig(d)?}

From the undoped case to a doping concentration of $-0.06$ $e$/f.u., the band inversion between bands 13 and 14 [red and blue lines in Fig.~\ref{allphonon}(a)] becomes stronger along $\Gamma$-M, pushing the violet square nodes along $\Gamma$-M closer to the M high-symmetry point.

Further increasing the doping concentration to $-0.08$ $e$/f.u. brings the two neighboring violet nodes together at M. At M, each neighboring pair of violet nodes carries the same frame charge, %either because the nodes already have the same frame charge, or 
because the nodes with opposite frame charges must cross the dark green Dirac string and the charge of one of the pair is flipped. Therefore, the pairs of violet nodes do not annihilate. Instead, they ``bounce'' to the M-K high-symmetry lines with each pair carrying the same frame charge, and we now label them as pink squares in Fig.~\ref{allphonon}(b) because their trajectories change. %In addition, one of the violet nodes in $\Delta_{13}$ crosses the dark green Dirac string in $\Delta_{14}$, and consequently one of the dark green nodes flips its sign as well. This flips the sign of the dark green nodes at K as well. 
The inversion between bands 14 and 15 also creates three pairs of same charged blue nodes and three pairs of same charged green nodes along $\Gamma$-M and K-$\Gamma$ respectively, indicating that the blue nodes must carry opposite frame charge with the green nodes so they can be created at the same time at $\Gamma$, or be annihilated simultaneously when brought back to $\Gamma$ by decreasing the doping concentrations from $-0.08$ to $-0.06$ $e$/f.u.

At $-0.10$ $e$/f.u. the nearest pairs of the blue nodes along $\Gamma$-M are brought together to M when the bands are fully inverted at M. During this process, one of the blue nodes in each pair must cross a Dirac string of the pink nodes and thus flips its sign. % both its sign and the sign of one pink node. 
Now that each nearest pair of the blue nodes near M has the opposite frame charge, they will be annihilated when meeting at M. As shown in Fig.~\ref{allphonon}(c), at $-0.10$ $e$/f.u. the blue nodes and their corresponding Dirac strings disappear. The disappearance of the blue Dirac string also flips the sign of the three pink nodes around the K point inside the Dirac string.

At $-0.12$ $e$/f.u., the inversion between bands 13 and 14 further increases, and as a result the pink nodes move further from M to K. Similarly, the green nodes move further from $\Gamma$ to K. During this process there is no conversion between the frame charges as no nodes meet together and no adjacent Dirac strings are crossed. 
As shown in Fig.~\ref{allphonon}(d), the global topological configuration is nearly the same, except that the pink and green nodes move closer to K.

The braiding process ends at $-0.14$ $e$/f.u., when all the bands are fully inverted and no further inversion occurs with increasing doping concentration. When the three pink nodes and three green nodes meet at K, there are two open pink squares, one closed pink square, and three closed green circles, as well as the open dark green circle at K. We can braid the two closed green nodes with one open pink node and one closed pink node so their charges are flipped. Eventually we will have one open pink square, two closed pink squares, one open green circle, two closed green circles, and one open dark green circle. There are two open and two closed circles in total, so the circles can be annihilated when brought together at K. The remaining open square and two closed squares can also recombine to form a closed square. Therefore we have a closed dark green square at K in the middle of the 2D Brillouin zone in Fig.~\ref{allphonon}(e). The braiding processes around other K points are similar.

\subsection{Topological edge states}

\begin{figure}
\centering
\includegraphics[width=0.9\linewidth]{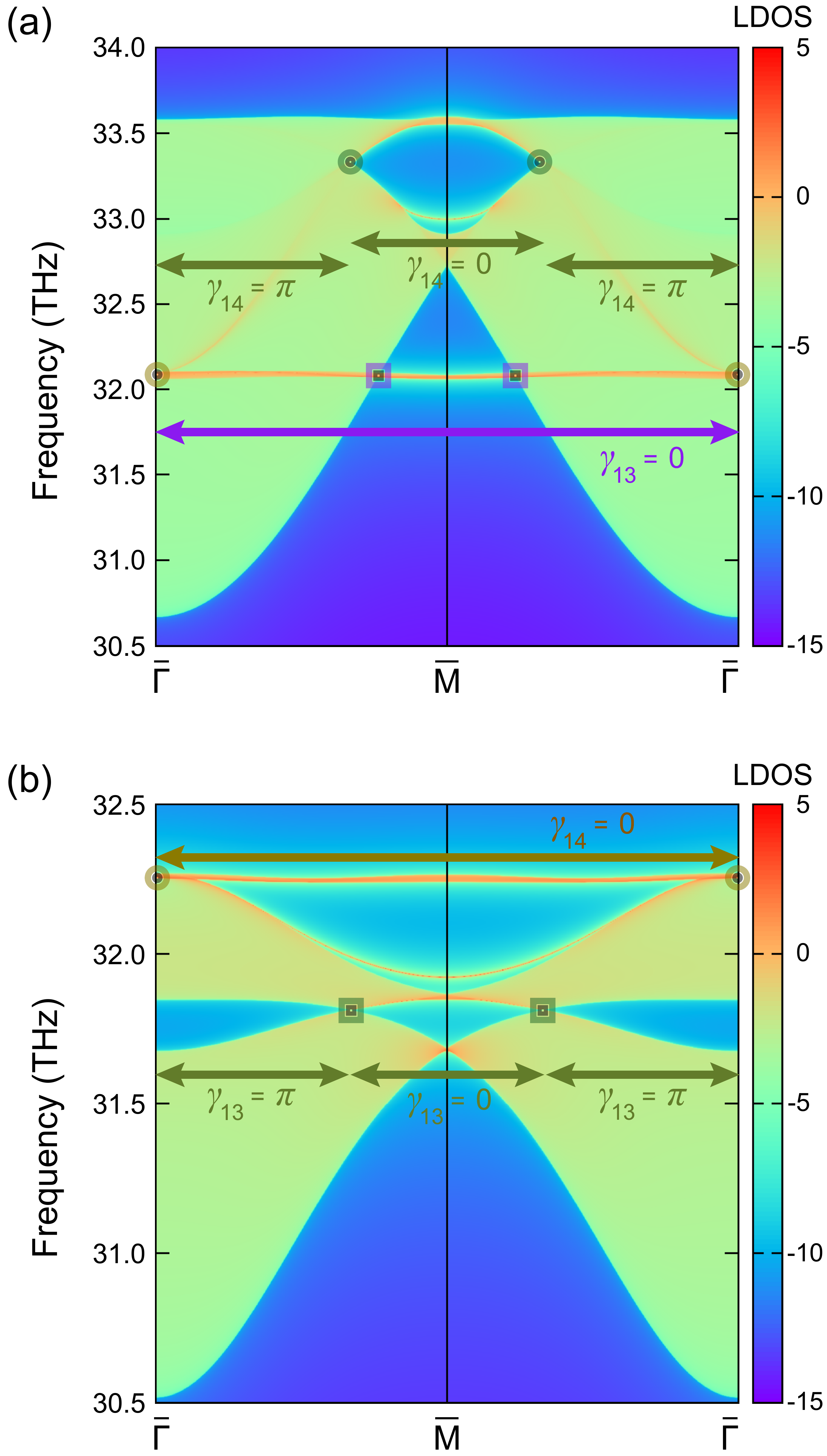}
\caption{Topological edge states along the (100) direction for (a) undoped and (b) $-0.20$ $e$/f.u. doped Al$_2$O$_3$.}
\label{edge} 
\end{figure}

Despite the fact that the bulk-boundary correspondence for multi-gap topologies has not yet been fully characterized (and is beyond the scope of this work), we can still investigate the evolution of the topological edge states of monolayer Al$_2$O$_3$ upon electrostatic doping. The surface local densities of states (LDOS) are calculated from the imaginary part of the surface Green¡¯s function as implemented in {\sc WannierTools} \cite{Wu2018}. We first compare the edge states for undoped and $-0.20$ $e$/f.u. doped Al$_2$O$_3$ in Fig.~\ref{edge}.

For undoped Al$_2$O$_3$, the edge states connecting the projections of a pair of bulk nodes in $\Delta_{14}$ at K (dark green circles) are clearly visible, as shown in Fig.~\ref{edge}(a). In addition, we can see the projections of the flat bulk band across the entire edge Brillouin zone, ending at the projections of a pair of bulk nodes in $\Delta_{14}$ at $\Gamma$ (dark yellow circles). The emergence of the edge states can be understood by computing the Zak phase $\gamma$ \cite{Zak1989}. In $\Delta_{13}$, we obtain $\gamma_{13} = 0$ along the entire edge Brillouin zone except at $\bar{\Gamma}$. In $\Delta_{14}$, $\gamma_{13} = 0$ corresponds to emerging edge states between the projections of dark green nodes at K, while $\gamma_{13} = \pi$ corresponds to vanishing edge states.

For $-0.20$ $e$/f.u. doped Al$_2$O$_3$, the flat band has the highest energy, and the projections of the pair of bulk nodes at $\Gamma$ (dark yellow circles) are also higher than the projections of other bulk nodes, as shown in Fig.~\ref{edge}(b). The edge states merge from the projections of dark yellow circles in $\Delta_{14}$ at $\Gamma$. The projections of the other pair of bulk nodes at K (dark green squares) are inverted to lower frequencies, redistributing their non-Abelian frame charges from $\Delta_{14}$ to $\Delta_{13}$. % Although the projections of the square nodes are well separated from other states, no edge states can be observed. 
Similar to the Zak phase in the undoped case, for $-0.20$ $e$/f.u. doped Al$_2$O$_3$, $\gamma_{13} = 0$ and $\gamma_{14} = 0$ correspond to emerging edge states, while $\gamma_{13} = \pi$ corresponds to vanishing edge states.

This is consistent with the fact that the oxygen atoms, as the atomic centers, are on the boundary of the unit cell. The Zak phase measures the displacement between the Wannier functions and the atomic centers. In our case, a Zak phase of zero indicates that the phonon Wannier functions occupy the center of the unit cell and are localized away from the atomic centers on the unit cell boundary, leading to an ``anomaly'' with localized edge states \cite{Jiang2021,Lange2021}. On the other hand, a Zak phase of $\pi$ indicates that the phonon Wannier states and the atoms are at the same places on the unit cell boundary, corresponding to vanishing edge states. While this $\mathbb{Z}_2$-quantized Berry phase is a good quantum number for 1D edge states \cite{Zak1989} and traces in {\it some} gaps the edge states faithfully, we repeat that a full bulk boundary relation describing topological phases obtained by non-Abelian processes is still subject to intense research activity.

%{\color{red} The oxygen atoms are in the 1/2 coordinates of the unit cell, so the Zak phase of 0 corresponds to emerging surface states. Do we need to mention it here?}

We also show the evolution of the edge states under phonon braiding upon electrostatic doping from $-0.06$ to $-0.14$ $e$/f.u. in the bottom panel of Fig.~\ref{allphonon}, which can provide information on the conversion of non-Abelian frame charges in the bulk states. Because the bulk nodes are distributed in a narrow frequency range between $31.8-32.8$ THz, the topological edge states are not well separated from each other.

%\subsection{Possible experimental signatures}

\section{Discussion}
\label{discussion}

Our findings of Section~\ref{case} suggest a broad relevance to the fields of topology, phonons, dielectrics, first principles modelling and information storage.

From a topological perspective, we find that phonons can be a primary platform to study multi-gap topologies. When studying multi-gap topologies in electronic systems, all three neighboring bands must be near the Fermi level, which severely limits the potential material candidates. On the other hand, phonons do not have the restriction of the Fermi level because they are bosonic excitations. In addition, the time reversal symmetry $T$ in phononic systems is hard to break, making it more convenient to find material candidates with $C_2T$ symmetry or $PT$ symmetry. As phonons can be treated as spinless systems, we can apply several existing models, such as the three-band spinless model with a Kagome lattice \cite{Jiang2021}, to phononic systems. We can also extend the ideas of non-Abelian braiding to other quasiparticles such as magnons \cite{Li2016d} and excitons \cite{Wu2017a}. An open question is the bulk-boundary correspondence of multi-gap topologies, and phonon dispersions with fewer, cleaner band crossings, e.g. phonon systems with only one or two atoms in the unit cell, could provide an ideal platform for its study. 

For the phonon community, we open a new research direction for these emergent excitations. Traditional studies of phonons mainly involve conventional superconductivity \cite{Drozdov2015,Ahadi2019}, electrical and thermal transport \cite{Peng2016,Peng2016g,Peng2018a,Peng2018b,Peng2019a}, carrier thermalization \cite{Bernardi2014}, structural phase transition \cite{Peng2020,Gu2021} and charge density waves \cite{Hill2019,Yue2020}. 
%the thermal transport properties of phonons upon external stimuli \cite{Wang2007,Zhou2019a}. 
Here we show that the reciprocal space braiding of phonon band crossing points can also form the basis for next-generation phononic computation. Several strategies could be used to experimentally verify the non-Abelian braiding of phonons, including inelastic neutron scattering \cite{Zhang2019c,He2020,Choudhury2008}, inelastic X-ray scattering \cite{Miao2018}, and high resolution electron energy loss spectroscopy \cite{Jia2017}. %The spectral features for different band orders are distinct, allowing for the experimental distinction between different topological configurations. 
Additionally, our first principles evaluation of the band inversion processes can provide references for the experimental observation of the evolution of the phonon band nodes upon electrostatic doping. Another open question is whether the band inversion at K can be measured by double resonance Raman modes \cite{Guo2015} (for possible scattering process, see Supplementary Material).

For dielectrics, we provide an experimentally realizable way to control the braiding in a well-known dielectric material upon electrostatic doping. Monolayer Al$_2$O$_3$ has been widely used as a gate dielectric in electronic devices \cite{Gilmer2002,Hirama2012,Ren2017,Kwon2020}. Therefore, electrostatic doping of monolayer Al$_2$O$_3$ can be experimentally feasible and has the potential to be incorporated into existing devices based on Al$_2$O$_3$, opening the door for studying topology-related phenomena in this otherwise well-studied material. In addition, electrostatic doping-induced phonon shifts have been intensively studied in low-dimensional materials, which can be probed directly using Raman spectroscopy \cite{Zhang2008,Sohier2019}. We would also like to highlight that applying an electric field of $0.65-0.95$ eV/\AA\ has similar effects on phonon band inversion and the corresponding braiding processes (for details, see Supplementary Material). It is therefore interesting to investigate how doping or gating redistributes the charge density and how the redistributed charge couples to the lattice vibrational modes.

For first principles modelling, we note that the computational techniques described in this article offer a route to understand non-Abelian braiding of \textit{any quasiparticle}. The Euler class can be computed using the eigenvectors/eigenstates of these quasiparticles as input. We offer a detailed description of all the theoretical background and computational methodology to analyse the non-Abelian frame charges formed by any three-band subsystems in the spectra of any quasiparticle, as long as the symmetry requirements are fulfilled. We anticipate that the potential braiding processes rely on control of band inversion and the corresponding redistribution of non-Abelian frame charge. Future simulation work could focus on various strategies to control the phonon braiding, including nonlinear effects \cite{Juraschek2017,Khalsa2018,Juraschek2021,Khalsa2021}, anharmonicity \cite{Borinaga2016,Boukhicha2013,Michel2015,Whalley2016,He2020}, and ultrafast pumps \cite{Ishioka2008,Reichardt2020,Trovatello2020,Karlsson2021}. 

Finally, as a further distant and more speculative perspective, the braid processes might find use in storing information. The idea is that information may be robust against perturbations from the environment because the non-trivial frame charges can only become trivial by unbraiding the non-Abelian frame charges via introducing a third phonon band. In addition, we can control these non-Abelian frame charges and their braiding by electrostatic doping, offering new opportunities for a conceptually new computation hardware based on phonons. Moreover, the braiding of multiple nodes takes place simultaneously when these nodes are related to each other by the space group symmetry, suggesting the possibility of storing information in the frame charges combining the topological and the symmetry information together. While we foresee that such information could be encoded in phonons, this nevertheless leaves open the exciting question of whether phonons could be suitable to implement quantum algorithms.
%If such information could be encoded in phonons, another open question would be whether we can implement a quantum algorithm suitable for phonons.

\begin{acknowledgements}
We thank Gunnar F. Lange for helpful discussions. B.P., R.-J.S., and B.M. acknowledge funding from the Winton Programme for the Physics of Sustainability. R.-J.~S. also acknowledges funding from the Marie Sk{\l}odowska-Curie programme under EC Grant No. 842901 and from Trinity College at the University of Cambridge. B.M. also acknowledges support from the Gianna Angelopoulos Programme for Science, Technology, and Innovation. The calculations were performed using resources provided by the Cambridge Tier-2 system, operated by the University of Cambridge Research Computing Service (www.hpc.cam.ac.uk) and funded by EPSRC Tier-2 capital grant EP/P020259/1, as well as with computational support from the U.K. Materials and Molecular Modelling Hub, which is partially funded by EPSRC (EP/P020194), for which access is obtained via the UKCP consortium and funded by EPSRC grant ref. EP/P022561/1.
\end{acknowledgements}

%\bibliography{new}


\end{document}